\documentclass[fleqn,usenatbib]{mnras}

\usepackage{newtxtext,newtxmath}

\usepackage[T1]{fontenc}
\usepackage{ae,aecompl}

\usepackage{graphicx}	
\usepackage{amsmath}	
\usepackage{amssymb}	

\usepackage[normalem]{ulem}

\newcommand{\kmsmpc}{\kms\;{\rm Mpc}^{-1}}

\newcommand{\hkpc}{h^{-1}{\rm kpc}}
\newcommand{\hmpc}{h^{-1}{\rm Mpc}}

\newcommand{\kms}{\;{\rm km}\,{\rm s}^{-1}}

\newcommand{\msolar}{\;{\rm M}_{\odot}}

\newcommand{\gad}{{\sc Gadget-3}}
\newcommand{\gizmo}{{\sc Gizmo}}

\newcommand{\simba}{{\sc Simba}}

\newcommand{\diff}{{\rm d}}

\usepackage{comment}

\title[]{The Dust-to-Gas and Dust-to-Metal Ratio in Galaxies from $z=0-6$}

\author[Q. Li, Narayanan \& Dav\'e]{Qi Li$^{1}$\thanks{pg3552@ufl.edu}, Desika Narayanan$^{1,2,3}$, \& Romeel Dav\'e$^{4,5,6}$\\$^{1}$Department of Astronomy,
  University of Florida, 211 Bryant Space Science Center, Gainesville,
  FL, 32611, USA\\$^{2}$University of Florida Informatics Institute, 432 Newell Drive, CISE Bldg E251 Gainesville,
  FL, 32611, US\\$^{3}$Cosmic Dawn Centre at the Niels Bohr Institue, University of Copenhagen and DTU-Space, Technical University of Denmark\\
  $^{4}$Institute for Astronomy, Royal Observatory, University of Edinburgh, Edinburgh, EH9 3HJ, UK\\$^{5}$University of the Western Cape, Bellville, Cape Town, 7535, South Africa\\
$^{6}$South African Astronomical Observatories, Observatory, Capte Town, 7925, South Africa}%

\date{}

\pubyear{2019}

\hypersetup{draft}

\begin{document}

\label{firstpage}
\pagerange{\pageref{firstpage}--\pageref{lastpage}}

\maketitle

\begin{abstract}
We present predictions for the evolution of the galaxy dust-to-gas (DGR) and dust-to-metal (DTM) ratios from $z=0\to 6$, using a model for the production, growth, and destruction of dust grains implemented into the \simba\ cosmological hydrodynamic galaxy formation simulation.  In our model, dust forms in stellar ejecta, grows by the accretion of metals, and is destroyed by thermal sputtering and supernovae.  Our simulation reproduces the observed dust mass function at $z=0$, but modestly under-predicts the mass function by $\sim\times 3$ at $z\sim 1-2$.  The $z=0$ DGR vs metallicity relationship shows a tight positive correlation for star-forming galaxies, while it is uncorrelated for quenched systems.  There is little evolution in the DGR-metallicity relationship between $z=0-6$.  We use machine learning techniques to search for the galaxy physical properties that best correlate with the DGR and DTM.  We find that the DGR is primarily correlated with the gas-phase metallicity, though  correlations with the depletion timescale, stellar mass and gas fraction are non-negligible.  We provide a crude fitting relationship for DGR and DTM vs. the gas-phase metallicity, along with a public code package that estimates the DGR and DTM given a set of galaxy physical properties.
\end{abstract}
\section{Introduction}

Dust plays a critical role in the physics of the interstellar medium (ISM) and galaxy evolution. The surfaces of dust grains catalyze a range of chemical reactions that influence the structure of ISM and star formation \citep{Hollenbach1971,Mathis1990,Weingartner2001,Draine2003,Wolfire2008,Hollenbach2012, Gong2017}, including the formation of molecular hydrogen and grain-catalyzed recombinations of H$^+$ and C$^+$. The ejection of dust from galaxies can contribute to metal abundances in the intergalactic medium and offers an additional cooling channel \citep[][]{Ostriker1973, Bouche2007,Peeples2014,Menard2010,Peek2015, Vogelsberger2018}, while dust absorption of far ultraviolet and optical photons can shape the temperature structure of the neutral ISM \citep{goldsmith01a,krumholz11a,narayanan11b,narayanan12a,narayanan12b}. 

A complex set of physical processes contributes to the evolving dust content of  the Universe. It can be produced via condensation of dust grains from the gas-phase metals in the ejecta of asymptotic giant branch (AGB) stars and supernovae (SNe; \citealt{Gehrz1989,Todini2001, Nozawa2003, Ferrarotti2006,Nozawa2007,Zhukovska2007,Nanni2013,Schneider2014}), after which it can grow in the ISM via accretion of gas-phase metals \citep{Dominik1997,Dwek1998,Hirashita2011,Zhukovska2014}. It can be destroyed via enhanced non-thermal sputtering in SN blast waves, thermal sputtering, and via grain-grain collisions \citep{Draine1979a,Draine1979b,Seab1983, McKee1987,Jones1996,Bianchi2005,Nozawa2007}. 

Dust properties in galaxies have been intensively studied through
statistics and scaling relations, of which three particularly
interesting are dust mass functions (DMFs)
\citep{Dunne2003,Vlahakis2005,Eales2009,Dunne2011,Clemens2013,Beeston2018},
dust-to-gas mass ratios (DGRs) and dust-to-metal mass ratios
(DTMs) as a function of galaxy metallicity or stellar mass
\citep{Issa1990,Lisenfeld1998,Hirashita2002,Draine2007,Galametz2011,Remy-Ruyer2014,Giannetti2017,Chiang2018,Kahre2018,DeVis2019,DeCia2013,DeCia2016,Zafar2013,Sparre2014,Wiseman2017}.
These relationships provide a convenient method for determining gas
masses in galaxies, as well as providing constraints on the baryon
cycle that governs galaxy evolution at low and high-redshifts \citep[e.g.][]{Magdis2012}.  

Theorists have commonly used an assumed constant dust-to-metal ratio in galaxies in order to model the evolving dust content in hydrodynamic or semi-analytic models of galaxy formation
\citep[e.g.][]{Silva1998,Granato2000,Baugh2005,Lacey2010,Narayanan2010,Narayanan2015,narayanan18a,narayanan18b,
Fontanot2011,Niemi2012,Somerville2012,Hayward2013,Cowley2017,Katz2019,Ma2019}.
However, there is growing evidence from both integrated and resolved far-infrared studies of galaxies at both low and high-redshift that the dust-to-gas and dust-to-metal ratios in galaxies are not constant, and may not even be straight-forwardly modeled by a simple linear relationship with a galaxy physical property (such as metallicity).  For example, while the gas to dust ratio appears to scale with the metallicity of galaxies in the local Universe \citep{Dwek1998,Draine2007,Bendo2010}, there may be deviations from this trend at the lowest metallicities \citep[e.g.][]{Galliano2005,Galametz2011,Remy-Ruyer2014,DeVis2019}.  Similarly, the DTG measured by damped Lyman-alpha (DLA) and gamma-ray burst (GRB)
absorbers \citep[e.g.][]{DeCia2013,DeCia2016,Wiseman2017} from $z=0.1$
to $z=6.3$ are similar to those in the Local group, though drop at
metallicities lower than 0.05~$Z_\odot$. Hence more sophisticated theoretical modeling of galaxy dust content and its evolution is needed. 

In recent years, galaxy evolution models have progressed from treating dust as a simple scale factor of the metal mass \citep[see][and references therein]{Somerville2012} to including the physics of dust formation, growth and destruction in galaxies as they evolve.  The first generation of these sorts of simulations treated galaxies as one-zone models \citep[e.g.][]{Issa1990,Dwek1998,Inoue2003,Morgan2003,Calura2008,Zhukovska2007,Hirashita2009,Asano2013,Calura2014,
Rowlands2014,Zhukovska2014,Feldmann2015,DeVis2017}, though more recently a number of groups have begun to incorporate self-consistent dust physics on-the-fly into bona fide hydrodynamic models of galaxy formation and evolution.  
\citet{Bekki2015}, \citet{McKinnon2016}, and \citet{Aoyama2017} established some of the initial frameworks for including dust in hydrodynamic galaxy formation simulations to study the evolution of dust 
properties in individual galaxies. Building on this, \citet{McKinnon2017} performed full-volume cosmological simulations 
using moving mesh code {\sc arepo} to study dust properties across galaxies over cosmic time, but were unable to successfully reproduce the DGR -- metallicity relation. \citet{Vogelsberger2018} extended this framework by implementing high temperature dust cooling channels to study dust in galaxy clusters and its impact on the intergalactic medium. \citet{Aoyama2017} and \citet{Aoyama2018} developed a $2$-grain size model into  a SPH cosmological simulation, where they studied overall dust properties in a whole cosmological volume and IGM, while \citet{Hou2019} built on this to add a phenomenological Active Galactic Nuclei  (AGN) feedback model. Finally, \citet{Popping2017} and \citet{Vijayan2019} have implemented the physics of dust formation, growth and destruction into semi-analytic galaxy formation models.  The growing interest in modeling dust evolution highlights its importance in more accurately modeling the observed properties of galaxies.

What has been missing thus far is a predictive self-consistent model for the dust-to-gas and dust-to-metal ratios in galaxies across cosmic time in a large-volume cosmological galaxy formation simulation.  In this paper, we aim to develop this model. To do this, we incorporate into state-of-the-art cosmological hydrodynamic simulation \simba\ \citep{Dave2019} a model to track  on-the-fly dust formation and evolution, broadly following the \citet{McKinnon2017} passive scalar dust algorithm.  Here, passive refers to the dust being advected with the gas, and scalar refers to the dust having a fixed grain size distribution.

We include dust production from Type II SNe and Asymptotic Giant Branch (AGB) stars, and further growth via accretion of metals, while destruction can occur from sputtering, consumption by star formation, or SN shocks. We explore the evolution of the galaxy dust mass function and the scaling relations of the DGR and DTM with metallicity over cosmic time. We then build on this, and investigate the physical drivers of the DGR and DTM using a machine-learning framework trained by our simulated dataset to understand the scatter in the DGR/DTM-metallicity relation. We use these tools to develop an algorithm (that we release publicly) for the dust mass from galaxies without the assumption of an overly simplistic dust-to-gas or dust-to-metal ratio. We additionally provide a simple scaling relation for the DGRs in galaxies.

This paper is organized as follows. In \S\ref{sec:method}, we summarize the \simba\ simulation suite, with a particular focus on the 
model for dust formation and evolution. We present the dust mass functions and scaling relations between the DGR/DTM and gas phase metallicities in  \S\ref{sec:cosmo}. In \S\ref{sec:dtg}, we model the underlying physical drivers of the DGR and DTM, and establish a connection between the DGR/DTM and various physical properties of galaxies. We then discuss our results, compare them to other theoretical work, and discuss potential caveats in \S\ref{sec:discussion}, and conclude  in \S\ref{sec:conclude}.

\section{Methodology}

\label{sec:method}
\subsection{Cosmological Simulations}
\label{sec:code}

This work utilizes the \simba\ cosmological hydrodynamic simulation . We refer the reader to \citet{Dave2019} for full details, and we summarize the salient points here.

The primary \simba\ simulation we use here has $1024^3$ dark matter particles and $1024^3$ gas elements in a cube of $100\hmpc$ side length, and is run from $z=249$ down to $z=0$. We assume a Planck16 \citep{Planck2016} concordant cosmology of $\Omega_m=0.3$, $\Omega_\Lambda=0.7$, $\Omega_b=0.048$, $H_0=68\kmsmpc$, $\sigma_8=0.82$, and $n_s=0.97$. Our \simba\ run has a minimum gravitational softening length $\epsilon_{\rm min} = 0.5 \hkpc$, mass resolution $9.6\times 10^7 \msolar$ for dark matter particles and $1.82\times 10^7 \msolar$ for gas elements. The system is evolved using a forked version of the \gizmo\ cosmological gravity plus hydrodynamics solver~\citep{Hopkins2015}, in its Meshless Finite Mass (MFM) version. This code, modified from \gad~\citep{Springel2005}, evolves dark matter and gas elements together including gravity and pressure forces, handling shocks via a Riemann solver with no artificial viscosity.

Radiative cooling and photoionisation heating are modeled using the {\sc Grackle-3.1} library~\citep{Smith2017}, including metal cooling and non-equilibrium evolution of primordial elements. An H$_2$-based star formation rate is used, where the H$_2$ fraction is computed based on the sub-grid model of \citet{Krumholz2009} based on the metallicity and local column density, with minor modifications as described in \citet{Dave2016} to account for variations in numerical resolution.  The star formation rate is given by the H$_2$ density divided by the dynamical time: SFR$=\epsilon_*\rho_{\rm H2}/t_{\rm dyn}$, where we use $\epsilon_*=0.02$~\citep{Kennicutt1998}. These stars drive winds in the interstellar medium. This form of feedback is modeled as a two-phase decoupled wind,  with 30\% of wind particles ejected hot, i.e. with a temperature set by the supernova energy minus the wind kinetic energy. The modeled winds have an ejection probability that scales with the the galaxy circular velocity and stellar mass (calculated on the fly via fast friends-of-friends galaxy identification). The nature of these scaling relations follow the results from higher-resolution studies in the Feedback In Realistic Environments zoom simulation campaign \citep[e.g.][]{Muratov2015,Angles2017b,Hopkins2014,Hopkins2018}.

The chemical enrichment model tracks eleven elements (H, He, C, N, O, Ne, Mg, Si, S, Ca, Fe) during the simulation, with enrichment tracked from Type II supernovae (SNe), Type Ia SNe, and Asymptotic Giant Branch (AGB) stars. The yield tables employed are: \citet{Nomoto2006} for SNII yields, \citet{Iwamoto1999} for SNIa yields, and AGB star enrichment following \citet{Oppenheimer2006}. Type Ia SNe and AGB wind heating are also included, along with ISM pressurisation at a minimum level as required to resolve the Jeans mass in star-forming gas as described in \citet{Dave2016}.

\simba\ incorporates black hole physics. Black holes are seeded and grown during the simulation via two-mode accretion. The first mode closely follows the torque-limited accretion model presented in \citet{Angles2017a}, and the second mode uses Bondi accretion, but solely from the hot gas component. The accretion energy is used to drive feedback that serves to quench galaxies, including a kinetic subgrid model for black hole feedback, along with X-ray energy feedback. \simba\ additionally includes a dust physics module to track the lifecycle of cosmic dust, which we describe in the following section.

\subsection{Modeling the Dust Lifecycle}
\label{sec:dust}

In our implementation, dust is fully coupled with gas flows. This treatment is essentially accurate, as the drift caused by the gas-dust drag force and the radiative pressure is under-resolved in our simulations. Additionally, dust grains are assumed to have the same physical properties with a constant radius $a\ =\ 0.1\ {\rm \mu m}$ and density $\sigma = 2.4\ {\rm g\ cm^{-3}}$ \cite{Draine2003}. We ignore dust cooling channels which will be implemented in future work.

Dust is produced by condensation of a fraction of metals from SNe and AGB ejecta. We follow the prescription described by Equation (4) to (7) in \citet{Popping2017} which updates the work of \cite{Dwek1998}. In the following, $m_{i,d}^j$ refers to the dust mass of the $i$th element (C, O, Mg, Si, S, Ca, Fe) produced by the $j$th stellar process (SNII or AGB stars), whereas $m_{i,{\rm ej}}^j$ refers to the mass of ejecta from the $j$th process.

The mass of dust produced by AGB stars with a carbon-to-oxygen mass ratio C/O $>$ 1 is expressed as
\begin{equation}
m_{i,d}^{\rm AGB}=
\begin{cases}
\delta_{\rm C}^{\rm AGB} (m_{C,{\rm ej}}^{\rm AGB} - 0.75 m_{O,{\rm ej}}^{\rm AGB}), & i\ =\ {\rm C}\\
0, & {\rm otherwise,}
\end{cases}
\end{equation}
where $\delta_i^{\rm AGB}$ is the condensation efficiency of element $i$ for AGB stars. The mass of dust produced 
by AGB stars with C/O $<$ 1 is expressed as
\begin{equation}
m_{i,d}^{\rm AGB}=
\begin{cases}
0, &  i\ =\ {\rm C}\\
16 \sum \limits_{i=\rm{Mg,Si,S,Ca,Fe}} \delta_i^{\rm AGB} m_{i, {\rm ej}}^{\rm AGB}, & i\ =\ {\rm O}\\
\delta_i^{\rm AGB} m_{i, {\rm ej}}^{\rm AGB}, & {\rm otherwise,}
\end{cases}
\label{eq:2}
\end{equation}
where $\mu_i$ is the mass of element $i$ in atomic mass units. The mass of dust produced by Type II SNe is described as
\begin{equation}
m_{i,d}^{\rm SNII}=
\begin{cases}
\delta_{\rm C}^{\rm SNII} m_{{\rm C}, {\rm ej}}^{\rm SNII}, &  i\ =\ {\rm C}\\
16 \sum \limits_{i=\rm{Mg,Si,S,Ca,Fe}} \delta_i^{\rm SNII} m_{i, {\rm ej}}^{\rm SNII}, & i\ =\ {\rm O}\\
\delta_i^{\rm SNII} m_{i, {\rm ej}}^{\rm SNII}, & {\rm otherwise,}
\end{cases}
\label{eq:3}
\end{equation}
where $\delta_i^{\rm SNII}$ is the condensation efficiency of element $i$ for SNII.

We choose a fixed dust condensation efficiency $\delta^{\rm AGB}_{i,\rm dust}=0.2$ based on the computation of \cite{Ferrarotti2006}, and $\delta^{\rm SN II}_{i,\rm dust}=0.15$ guided by \citet{Bianchi2007} to match the low-metallicity end of the observed $z=0$ relation between the DGR and gas-phase metallicities \citep{Remy-Ruyer2014}\footnote{Note - because our condensation efficiency for Type II SNe is tuned to match the low-$Z$ end of the local DGR-metallicity relation, the $z=0$ version of this relationship in the remainder of this paper should be treated as matching observations by construction, and not as a bona fide prediction.}. We omit the condensation of Type Ia SNe ejecta, as recent work suggests that Type Ia SNe are not significant sources of dust production \citep[see][]{Nozawa2011,Dwek2016,Gioannini2017}. This is different from \citet{McKinnon2016} and \citet{Popping2017} where Type Ia SNe are assumed to have the same condensation efficiency as Type II SNe.

Once dust grains are seeded, they grow by accreting gas-phase metals. Following \cite{Dwek1998}, the growth rate is expressed as
\begin{equation}
\left( \frac{\diff M_d}{\diff t} \right)_{\rm grow}=\left( 1-\frac{M_d}{M_{\rm metal}} \right) {\left( \frac{M_d}{\tau_{\rm accr}} \right)},
\end{equation}
where $M_{\rm metal}$ is the total mass of dust and local gas-phase metals. Following \citet{Hirashita2000} and \citet{Asano2013}, the accretion time scale $\tau_{\rm accr}$ is 
\begin{equation}
\tau_{\rm accr} = \tau_{\rm ref} \left( \frac{\rho_{\rm ref}}{\rho_g} \right) \left( \frac{T_{\rm ref}}{T_g} \right) {\left(\frac{Z_\odot}{Z_g} \right)}.
\label{eq:tauaccr}
\end{equation}
where $\rho_g$, $T_g$ and $Z_g$ are the local gas density, temperature and metallicity, respectively. 
$\rho_{\rm ref}$, $T_{\rm ref}$ and $Z_{\rm ref}$ are the reference values correspondingly. We take $\rho_{\rm ref} = 100$~H~atoms~cm$^{-3}$, $T_{\rm ref}=20$~K and $\tau_{\rm ref} = 10$~Myr in this work.

Dust grains can be eroded by colliding with thermally excited gas especially in hot halos (e.g. \citealt{Barlow1978,Draine1979b,Tielens1994}). We adopt the approximation of the thermal sputtering rate of grain radii derived by \cite{Tsai1995}, following \citealt{McKinnon2017} and \citealt{Popping2017}. The sputtering time scale is expressed as
\begin{equation}
\begin{aligned}
\tau_{\rm sp} & = a \left| \frac{\diff a}{\diff t} \right|^{-1}\\ 
 &\sim (0.17{\rm Gyr})\left( \frac{a}{0.1 \mu m} \right) \left( \frac{10^{-27}{\rm g\ cm^{-3}}}{\rho_g} \right)\left[ \left( \frac{T_0}{T_g}\right)^{\omega}+1 \right],
\end{aligned}
\end{equation} 
where $\omega$~=~$2.5$ controls the low-temperature scaling of the sputtering rate and 
$T_0\ =\ 2 \times 10^6$~K is the temperature above which the sputtering rate flattens. The growth rate of dust mass due to thermal sputtering is then calculated by
\begin{equation}
\left( \frac{\diff M_d}{\diff t} \right)_{\rm sp} = -\frac{M_d}{\tau_{\rm sp}/3}
\end{equation}

Because SN blast waves are not resolved in our simulations, we implement a subgrid model for dust destruction by SN shocks \citep{Dwek1980,Seab1983, McKee1987,McKee1989}. The characteristic time scale $\tau_{\rm de}$ is
\begin{equation}
\tau_{\rm de} = \frac{M_g}{\epsilon \gamma M_s},
\end{equation}
where $M_g$ is the local gas mass, $\epsilon = 0.3$ is the efficiency with which grains are destroyed in SNII shocks \citep{McKee1989}, $\gamma$ is the local SNII rate, and $M_s$ is the mass of local gas shocked to at least 100~km/s, calculated using the Sedov-Taylor solution to a homogeneous medium of $n_{\rm H}=0.13$~H~atoms~cm$^{-3}$ (the minimum SF threshold density of our simulations).

We additionally destroy dust completely in hot winds and during star formation and AGN X-ray heating (\S\ref{sec:code}). The parameters adopted in this simulation is listed in Table~\ref{tab:1}.

Finally, we note that for the star formation and grain growth models, we need to provide a total metallicity in solar units.  For this, we assume a solar abundance ($Z_\odot = 0.0134$) taken from \cite{Asplund2009}.

\begin{table*}
	\centering
	\caption{Simulation Free Parameters\label{tab:1}}
	\begin{tabular}{lccr} 
		\hline
		Parameter & Description & Value & Range accepted by literatures\\
		\hline
		Thermal sputtering & & & \\
		$a$ & Grain radius ($\mu m$) & 0.1 & ---\\
		$\sigma$ & Density of solid matters within grains (g cm$^{-3}$) & 2.4 & 2.2(graphite), 3.3(silicate)$^a$ \\
		Production & & & \\
		$\delta^{\rm AGB,C/O>1}_{i,\rm dust}$ & Condensation efficiency & 0.2 for $i$ = C  & 0.2 -- 1.0$^b$ \\
		                                     &                        &  0 otherwise      & 0       \\
		$\delta^{\rm AGB,C/O<1}_{i,\rm dust}$ &                       & 0 for $i$ = O  & 0\\
		                                      &                       &  0.2 otherwise  & 0.2 -- 0.8$^b$    \\
		$\delta^{\rm SNII}_{i,\rm dust}$     &                         & 0.15 for $i$ = C  & 0.15 -- 1.0$^b$\\
		                                      &                       &  0.15 otherwise  & 0.15 -- 0.8$^b$ \\       
		Growth & & &\\
		$\rho^{\rm ref}$ & Reference density (g cm$^{-3}$) & $2.3\times 10^{-22}$ & --- \\
		$T^{\rm ref}$ & Reference temperature (K) & 20 & --- \\
		$\tau_{\rm g}^{\rm ref}$ & Growth time-scale with $T=T^{\rm ref}$ and $\rho = \rho^{\rm ref}$ (Myr)   & 10 & 2 -- 500$^c$   \\
		Destruction (SNe Shock) & & &\\
		$E_{\rm SN,51}$ & Energy per SN ($10^{51}$ erg) & 1.0 & ---  \\
		$\epsilon$ & The efficiency of destruction by SN shocks & 0.3 & 0.1 -- 0.5$^d$\\
		\hline
	\end{tabular}
	\\$^a$ See \citet{Jones1996}.
	\\$^b$ See \citet{Dwek1998,McKinnon2017,Popping2017}
    \\$^c$ We fix $\rho^{\rm ref}=2.3\times 10^{-22}$~g~cm$^{-3}$ and $T^{\rm ref}=20$~K. See \citet{Dwek1998,Zhukovska2014,McKinnon2017,Popping2017}.
    \\$^d$ See \citet{McKee1989}.
\end{table*}

\subsection{Data Analysis -- A Machine-Learning Framework}
\label{sec:ert}

We seek to accurately quantify how galaxy dust properties, particularly the DGR, trace other global galaxy properties.  This represents a regression problem, where from a set of input variables, the prediction for the DGR is desired that most closely follows what is predicted directly by the simulation.

We employ machine learning for this regressor, as is now becoming common for a wide variety of astrophysical applications \citep[e.g.][]{Ball2007,Fiorentin2007,Gerdes2010,Kind2013,Ness2015,Kamdar2016,Agarwal2018,Rafieferantsoa2019}. Taking advantage of the large training set offered by \simba\ simulation of tens of thousands of galaxies, we use machine learning to relate the galaxy DGR to a set of galaxy properties, i.e. an $N$-dimensional vector {\boldmath $X$}, the components of which are the global galaxy properties as detailed in \S\ref{sec:dtg}.

The primary algorithm used in this work is extremely randomized trees (ERTs; \citealt{Geurts2006}). ERTs build a large ensemble of regression trees, each of which splits the training set -- here, an $(N+1)$-dimensional space comprising of data points ({\boldmath $X$},DGR) from 70\% of the simulated galaxies -- recursively among one randomly-selected subset of the galaxy properties. Each splitting divides the $(N+1)$-dimensional space into two $(N+1)$-dimensional subspaces, and it stops once the resulting subspace only contains one ({\boldmath $X$},DGR) point or the user-defined maximum tree depth is reached, in which case a relation between {\boldmath $X$} and DGR is established. The estimates produced by all the regression trees in the ERT ensemble are averaged to build a final map from {\boldmath $X$} to DGR. We refer readers interested in further details to  \citealt{Geurts2006} for the details of splitting and randomization in ERTs. For this work, we used the implementation of ERTs in the {\sc python} package, {\sc scikit-learn} \citep{Pedregosa2011}.

\section{Dust Properties Over Cosmic Time}
\label{sec:cosmo}

\subsection{Dust Mass Functions}

\begin{figure}
\includegraphics[width=0.5 \textwidth]{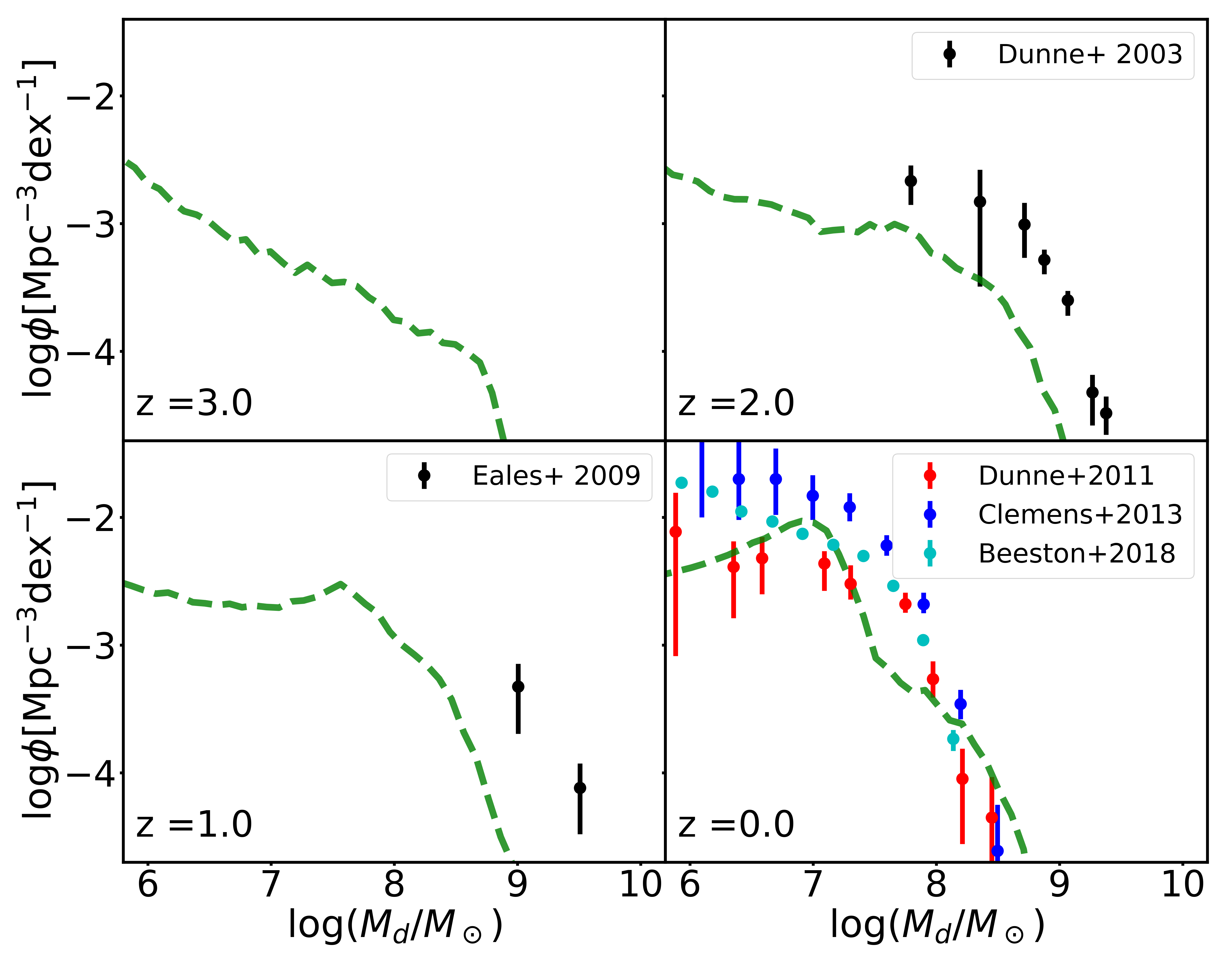}
\caption{Model dust mass functions from our cosmological galaxy formation simulations at redshifts $z=0-3$ for the full cosmological box, compared against observed data. When comparing to observational data sets, we select galaxies within particular redshift bins as follows. For \citet{Eales2009}, we plot data from $0.6 < z < 1.0$. For \citet{Dunne2011} and \citet{Beeston2018}, we plot data from $0.0 < z < 0.1$.  We standardized their results to our cosmological parameters (c.f. \S\ref{sec:code}) and the dust mass absorption coefficient $\kappa(850\ \mu {\rm m}) = 0.77\ {\rm cm^2\ g^{-1}}$.
\label{fig:dmf}}
\end{figure}

Figure~\ref{fig:dmf} shows the redshift evolution of dust mass function (DMF), comparing against the observational result of \citet{Dunne2003} at $z=2$, \cite{Eales2009} at $z = 1$, and \cite{Dunne2011,Clemens2013} and \cite{Beeston2018} at $z=0$. Unlike the comparison presented in \citet{Dave2019}, here we standardized their results to our cosmological parameters (c.f. \S\ref{sec:code}) and our assumed dust mass absorption coefficient $\kappa(850\ \mu {\rm m}) = 0.77\ {\rm cm^2\ g^{-1}}$. 

At $z=0$, \simba\ agrees well with observed data. Our simulation underproduces the DMF at the low mass end, due to our mass resolution and the minimum mass of identified galaxies (24 baryonic particles $\approx 4.37\times 10^8 \msolar$ baryonic mass).  The $z=2$ model dust mass function under-predicts the observational one by a modest factor $\sim 3$. This is still much better than early attempts in this area, where galaxies with $M_d \gtrsim 10^{8} M_\odot$ are hardly produced \citep[e.g.][]{McKinnon2017}.  We note that the observational mass function by \citet{Dunne2003} and \citet{Dunne2011} are from surveys of sub-mm sources with large beam sizes, which could result in multiple objects being blended within one beam therefore overestimating their dust masses \citep[e.g.][]{Narayanan2010,Hayward2013,Narayanan2015}. Beyond this, once we fold in the uncertainties in deriving dust masses from sub-millimetre photometry, it is probably premature to use the high-redshift DMF as a strong constraint on models. Overall, the dust mass function predicted by \simba\ broadly agrees with currently observed determinations, with very good agreement in the overall dust mass function shape.  This indicates that \simba\ viably models dust evolution over cosmic time in galaxies, and sets the stage for examining more detailed dust-related properties.

\begin{figure}
\includegraphics[width=0.5 \textwidth]{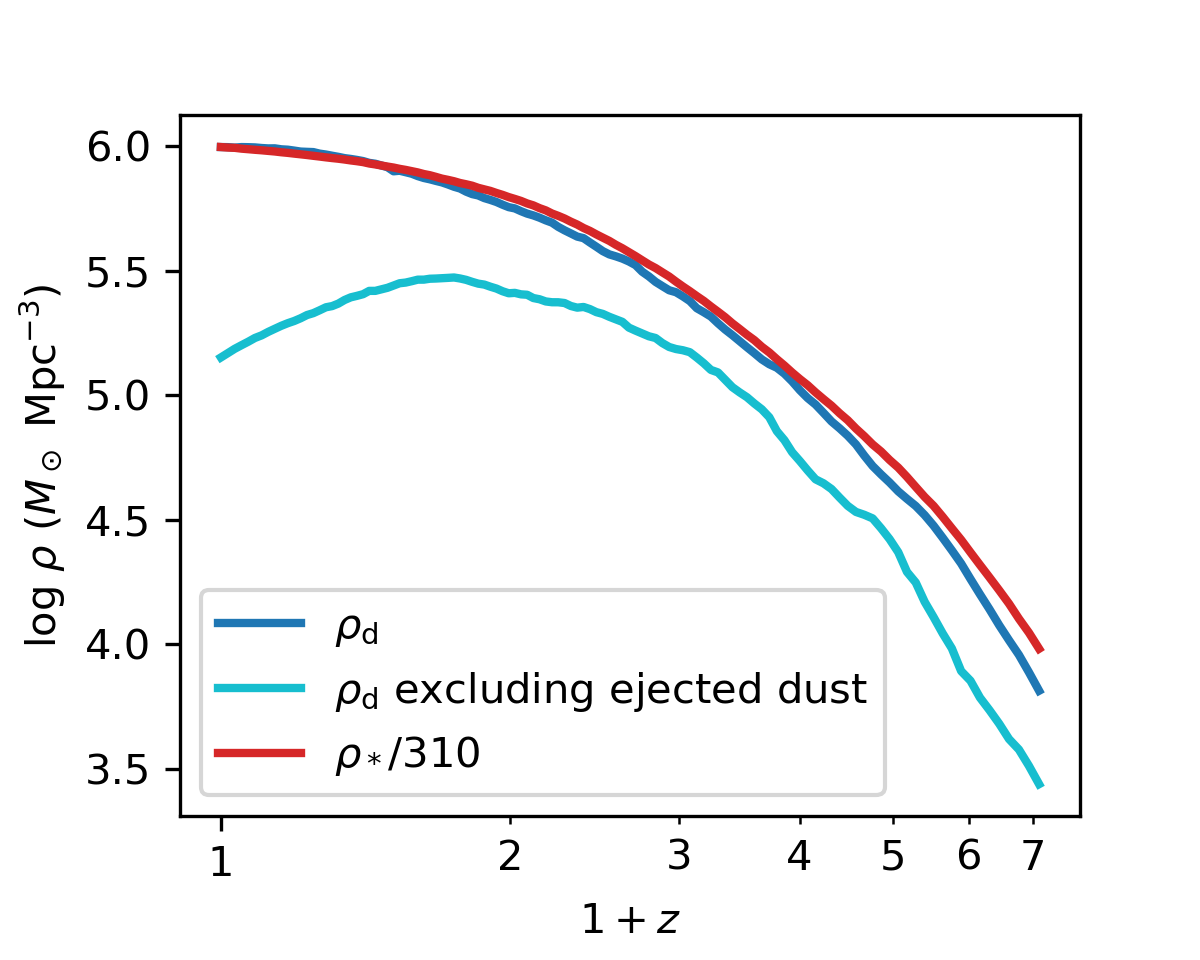}
\caption{The comoving cosmic dust mass density $\rho_{\rm d}$ (blue), the comoving dust mass density excluding dust ejected out of galaxies via galactic winds (cyan) and the comoving cosmic stellar mass density (red) $\rho_{*}$ as a function of redshift. $\rho_{*} = 310 \rho_{\rm d}$ at $z=0$. For the convenience of comparison, $\rho_{*}$ is normalized such that $\rho_{*}=  \rho_{\rm d}$ at $z=0$.} \label{fig:rhod}
\end{figure}

Figure~\ref{fig:rhod} shows the ratio of the comoving cosmic dust mass density $\rho_{\rm d}$ and the comoving cosmic stellar mass density $\rho_{*}$ as a function of redshift. The cosmic dust (or stellar) mass density is computed by summing the dust masses $M_d$ (or stellar masses $M_*$) of all gas cells and dividing by the total comoving volume. We get $\rho_{*} = 310 \rho_{\rm d}$ at $z=0$. For the convenience of comparison, $\rho_{*}$ is normalized such that $\rho_{*}= \rho_{\rm d}$ at $z=0$.

The cosmic dust mass density rapidly increases from $z=6$ by over 1.5 dex.  At late times, the dust density flattens as the global star formation rate falls, the amount of metals available to be accreted drops, and a quasi-balance is reached among dust production, growth, destruction, and astration.  The evolution of the dust mass density only slightly lags behind the stellar mass density at early epochs and catches up with rapid grain growth.

The comoving dust mass density excluding dust ejected out of galaxies via galactic winds is shown as the cyan line. It increases monotonically at high redshifts, following the total dust mass density. At $z\sim2$, it starts declining as star formation rates decline on average with the onset of quenching massive galaxies, which slows down the metal enrichment and thus limits the grain growth.  Meanwhile the destructive processes remain strong, and are even enhanced in massive galaxies that harbor little cold gas. Comparing this trend to the evolution of total dust mass density, we infer that the destruction of dust ejected to halos, dominated by thermal sputtering, is not strong enough to quickly eliminate dust grains owing to the low gas density. The wind model may need to be modified so that dust can be efficiently destroyed during the decoupled wind phase from galaxies into circum-galactic gas.
\subsection{The Dust-to-Gas Ratio}
\label{sec:dtgz}

\begin{figure}
\includegraphics[width=0.5 \textwidth]{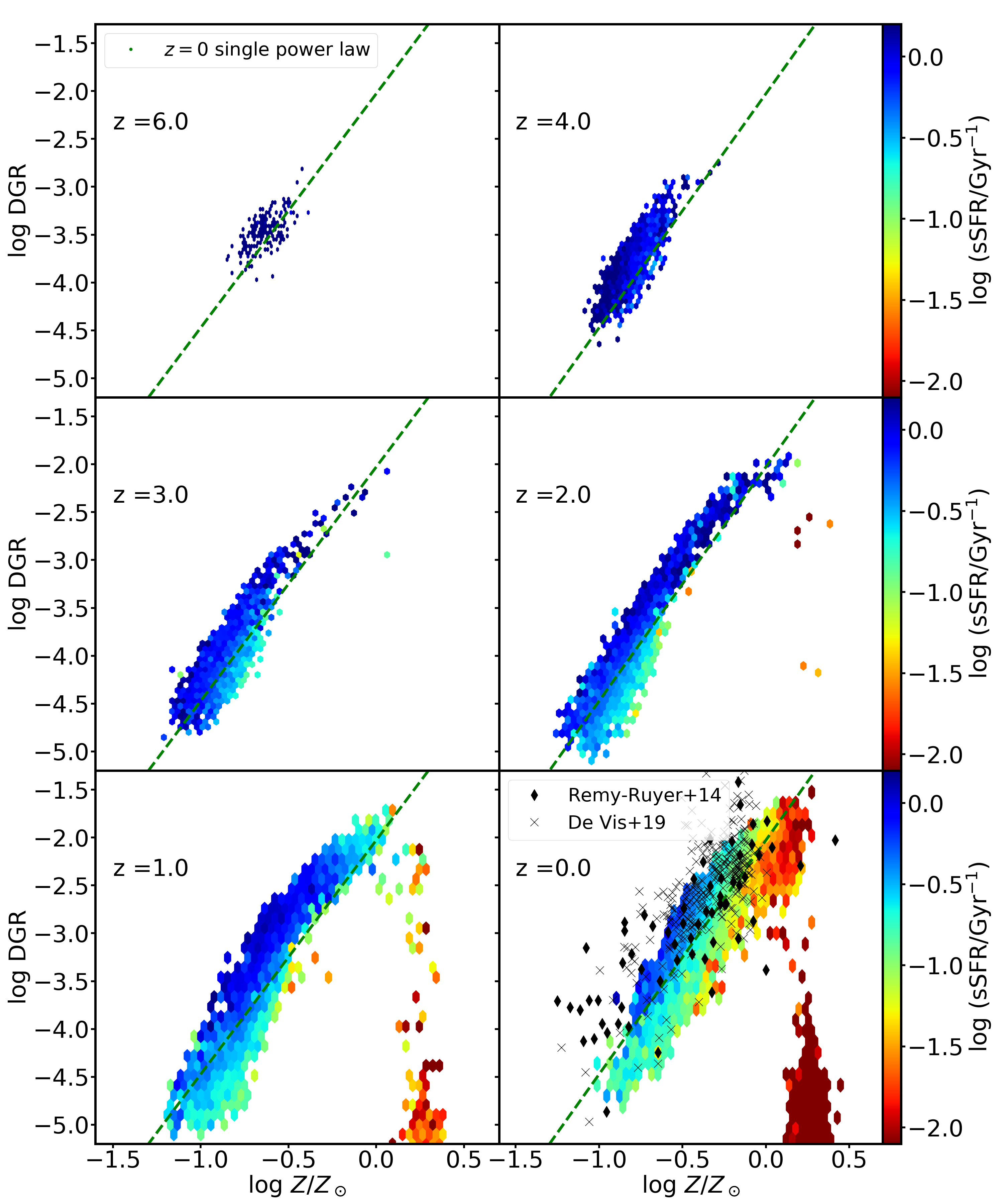}
\caption{The relation between dust-to-gas ratio (DGR) and gas-phase metallicity at $z$~=0-6. For $z$~=1-6, the theoretical data is shown with hexbins color-coded with specific star formation rates, while the best fit of $z=0$ theoretical relation with a power law (Equation~\ref{equation:DGR_fit}) is shown in each panel (for reference) with green dashed lines. The black dots and crosses in each panel are the $z=0$ observational data by \citet{Remy-Ruyer2014} and \citet{DeVis2019}, respectively. For \citet{DeVis2019}, we use metallicities derived from ``$S$" calibration of \citet{Pilyugin2016}.\label{fig:dtgz}}
\end{figure}

The agreement of our predicted dust mass functions with observations suggests that \simba\ represents a plausible dust evolution model. We now turn to examining the main target relations of our paper, the dust-to-gas and dust-to-metal ratios in \simba\ galaxies, in comparison to data.

Figure~\ref{fig:dtgz} shows the hexbin plot of the DGR as a function of gas-phase metallicity ($Z_{\rm gas}$) between $z=0-6$, compared against the data points as observed by \citet{Remy-Ruyer2014} and \citet{DeVis2019}. Hexbins are color-coded with specific star formation rates $\text{sSFR} = \text{SFR}/M_*$.

At $z=0$, \simba\ shows a DGR that is in good agreement with observations. However, we emphasise that our dust model was tuned to do so via our choice of the dust condensation efficiencies.  These quantities mostly change the $Z<0.2Z_\odot$ part of the DGR, without changing the slope much.  Hence the slope of the DGR vs. $Z_{\rm gas}$ is a robust prediction, as is our predictions for the redshift evolution.

There are two main regimes in the DGR$-Z_{\rm gas}$ plane.  The first regime corresponds to star-forming galaxies, where the DGR increases with $Z_{\rm gas}$. Our models predict a weak evolution of the DGR-$Z_{\rm gas}$ relation between $z=0$ to $z=6$, as is also predicted by \cite{Popping2017} using a semi-analytic model. The evolution of this relation in the DGR--$Z_{\rm gas}$ plane is mainly driven by the metal enrichment of galaxies as more galaxies just move along the sequence to slightly higher $Z_{\rm gas}$ at lower redshifts.
The second regime corresponds primarily to quenched galaxies, and shows low DGR values and no correlation with $Z_{\rm gas}$.
This is driven by AGN feedback that builds up the quenched galaxy population, in which dust production is stopped but dust destruction is enhanced by sputtering in surrounding hot gas. 
Even though there are no observed galaxies in this regime, this is likely due to an observational selection effects rather than an actual lack of galaxies \citep{DeVis2019}.
Finally, we note that the predicted DGR ratios in low-$Z_{\rm gas}$ galaxies generally lie below the observations.  These objects are typically very gas-rich dwarfs.  It is unclear whether they are overly gas-rich in \simba, or else they have too little dust production (or both). There may also be observational selection effects that bias in favor of higher dust masses in such small, faint systems.

We determine a best-fit power law to the DGR-$Z_{\rm gas}$ relation in the star-forming regime.  We separate the star-forming sequence from quenched galaxies by applying a density-based spatial clustering of applications with noise (DBSCAN) algorithm \citep{Ester1996} to galaxies in \{DGR, $Z_{\rm gas}$, sSFR\} space.  We then fit the star-forming sequence in the DGR--$Z_{\rm gas}$ plane using a power law:
\begin{equation}
\label{equation:DGR_fit}
\log \text{DGR} = (2.445 \pm 0.006)\log \left( \frac{Z}{Z_\odot} \right) - (2.029 \pm 0.003).
\end{equation}

This is quite close to the best fit power law to the \citet{DeVis2019}'s data $\log \text{DGR} = (2.45 \pm 0.12) \log \left( \frac{Z}{Z_\odot} \right) - (2.0 \pm 1.4)$, quantitatively confirming the good agreement of the predicted and observed slopes.  Still, the scatter of the best fit is large ($\sigma = 0.31$~dex), even though the correlation is clear. Better estimates of the DGR might be obtained by incorporating secondary physical parameters in addition to $Z_{\rm gas}$; we explore this in \S\ref{sec:dtg} using machine learning.

\begin{figure}
\includegraphics[width=0.5 \textwidth]{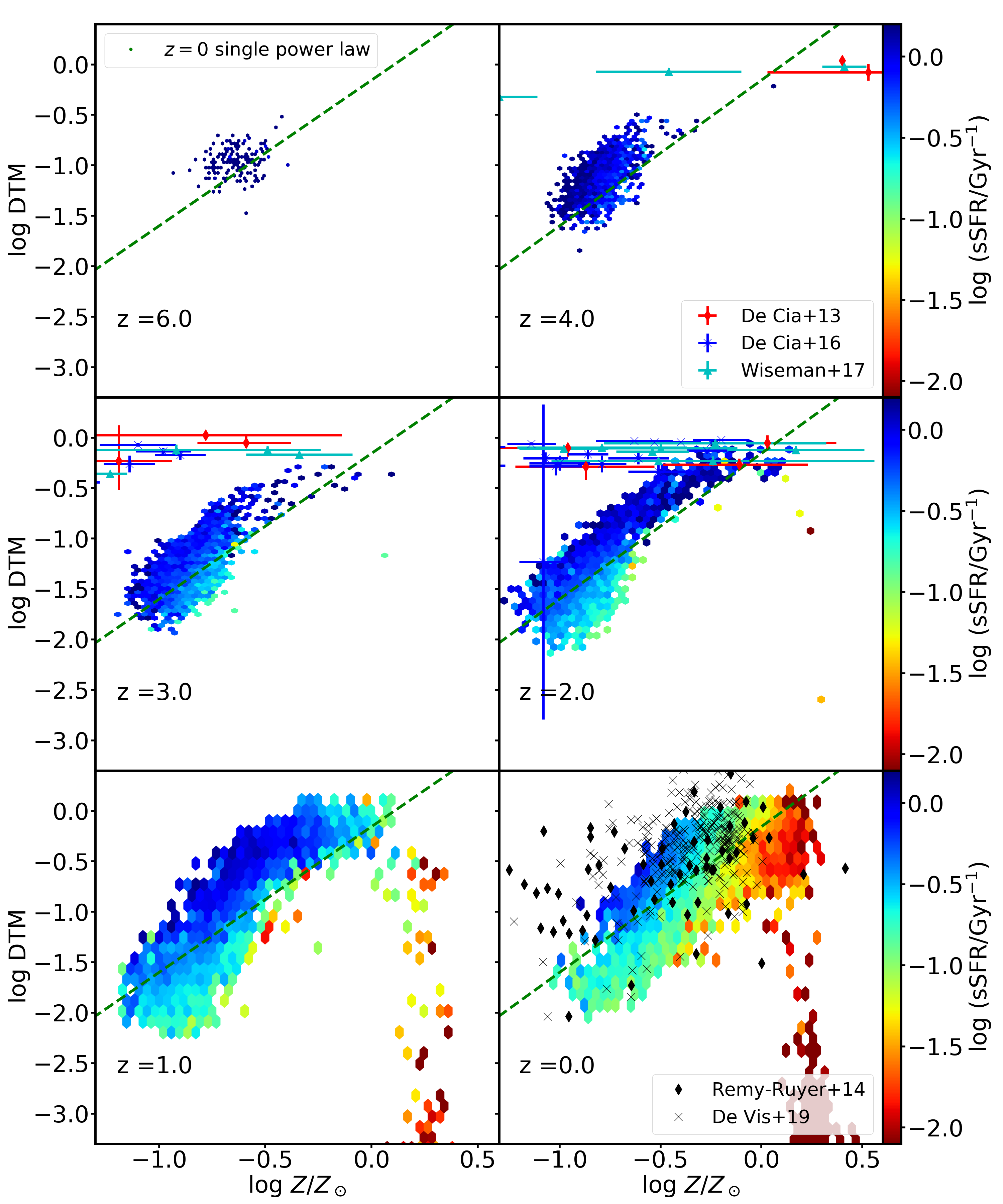}
\caption{The relation between dust-to-metal ratio (DTM) and gas-phase metallicity at $z$~=0-6. Observational data from high redshift observations of DLA and GRB absorbers \citep{DeCia2013,DeCia2016,Wiseman2017} are overplotted. The black dots and crosses in each panel are the $z=0$ observational data by \citet{Remy-Ruyer2014} and \citet{DeVis2019}, respectively. \label{fig:dtmz}}
\end{figure}

\subsection{The Dust-to-Metal Ratio}

In Figure~\ref{fig:dtmz}, we apply a similar analysis to the DTM, plotting the DTM ratio as a function of $Z_{\rm gas}$ between $z=0-6$. This relation is equivalent to the relation between the dust-to-gas ratio and $Z_{\rm gas}$. As discussed above, our simulation shows only a weak evolution of the DTM ratio from $z=6$ to $z=0$ and approximately reproduces the result of \citet{Remy-Ruyer2014} and \citet{DeVis2019}. The DTM ratio increases rapidly as $Z_{\rm gas}$ increases at the low metallicity regime until it is capped at a roughly constant value $\sim 0.8$ when $Z_{\rm gas} > 0.5 Z_{\odot}$.   For the quenched galaxies, the DTM ratio drops off quickly for the same reasons as in the DGR case.

Overplotting the $z\sim 2-4$ data from high redshift observations of Damped Lyman Alpha (DLA) and Gamma Ray Burst (GRB) absorbers \citep{DeCia2013,DeCia2016,Wiseman2017} against the simulated data, however, we find a systematic discrepancy at $Z_{\rm gas} < 0.5 Z_{\odot}$, where the predicted DTM ratios show a much steeper dependence with metallicity than the observations.
The source of this discrepancy is unclear.  We note that the nature of these absorbers can be significantly different from the physical conditions in galaxy disks. DLAs are thought to arise from the outskirts of gas disks in galaxies and perhaps even from metal-poor gas in the circum-galactic medium \citep{Berry2014}.  Moreover, these studies measure metallicities and DTM via abundances acquired from optical/UV absorption-line spectroscopy, which is different from methods typically used for galaxies observed in the local universe (i.e. strong line calibrations for metallicities and infrared emission for the dust mass).  This discrepancy thus may reflect a difference in the dust production versus destruction in different environments or scales, combined with the selection effect and methodologies of observations. We defer a more careful comparison of the DGR in these particular types of objects to future work, but note that there is potentially a discrepancy at low-$Z$.  This also may be responsible for the too-low DGR at low-$Z$.  We note, however, that significantly larger amounts of dust in low-mass galaxies would steepen the dust mass function, which may put our currently viable predictions into conflict with observations.

\subsection{Better DGR and DTM prediction via machine learning}
\label{sec:dtg}

We now investigate the physical drivers of the DGR and DTM ratios in galaxies, as well as their scatter.  While equation~\ref{equation:DGR_fit} provides a rough fit, it is clear that there is correlated scatter, and hence fitting with more variables should give a tighter relation.  To approach this agnostically in terms of the form and input variables, we employ a machine learning approach using Extremely Randomized Trees (ERTs).

We seek to establish a map between a range of physical properties -- namely gas-phase metallicity ($Z_{\rm gas}$), gas depletion time scale ($\tau_{\rm depletion}\equiv M_g/{\rm SFR}$), stellar mass ($M_*$), half-baryonic mass radius ($R_{b,50}$), gas mass fraction ($f_g\equiv M_g/(M_g+M_*)$), and gas surface density ($\Sigma_g$) -- onto the galaxy DGR and DTM ratios.  We limit our analyses to $z=0$ and concentrate our efforts on fitting the star-forming sequence of galaxies, since quenched galaxies show little dust and no obvious trend with any physical property and our relations show little evolution with redshift.  It would be straightforward to apply this methodology to other redshift outputs.

\begin{figure}
\includegraphics[width=0.5 \textwidth]{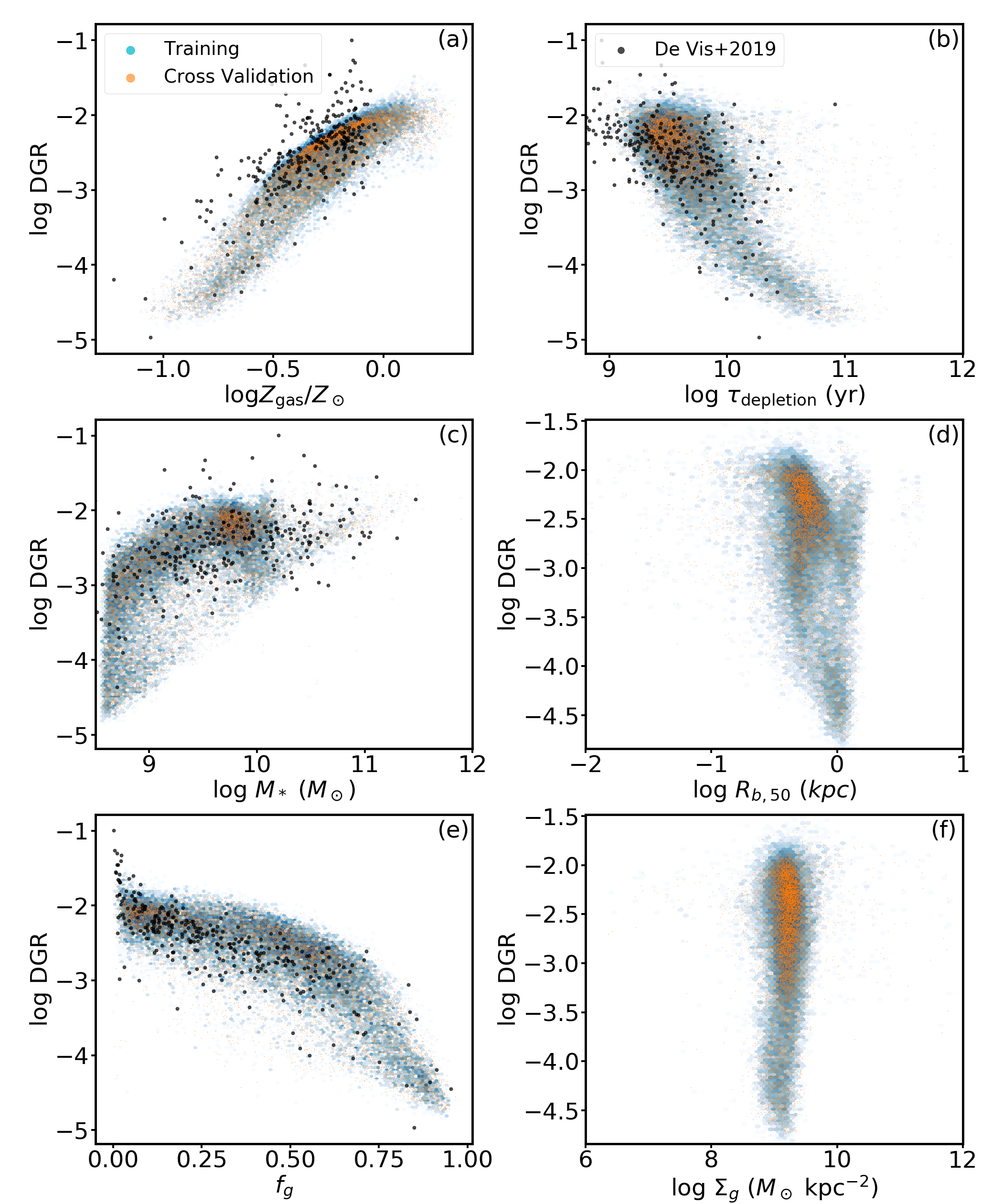}
\caption{The result our best-fit map from galaxy properties (gas-phase metallicity $Z_{\rm gas}$, gas depletion time scale $\tau_{\rm depletion}$ , stellar mass $M_*$, half-baryonic mass radius $R_{b,50}$, gas mass fraction $f_g$, and gas surface density $\Sigma_g$) to the DGR at $z=0$, using extreme randomized trees (ERT). The ERT is trained with the training set, denoted by cyan points, which consists of 70\% randomly-selected star-forming galaxies from \simba.  Orange points denote the prediction using galaxy properties of the cross-validation set which consists of the remaining 30\% of the galaxies. The black dots represent the $z \sim 0$ observational data by \citet{DeVis2019} for reference.
\label{fig:dtg_fit}}
\end{figure}

\begin{figure}
\includegraphics[width=0.5 \textwidth]{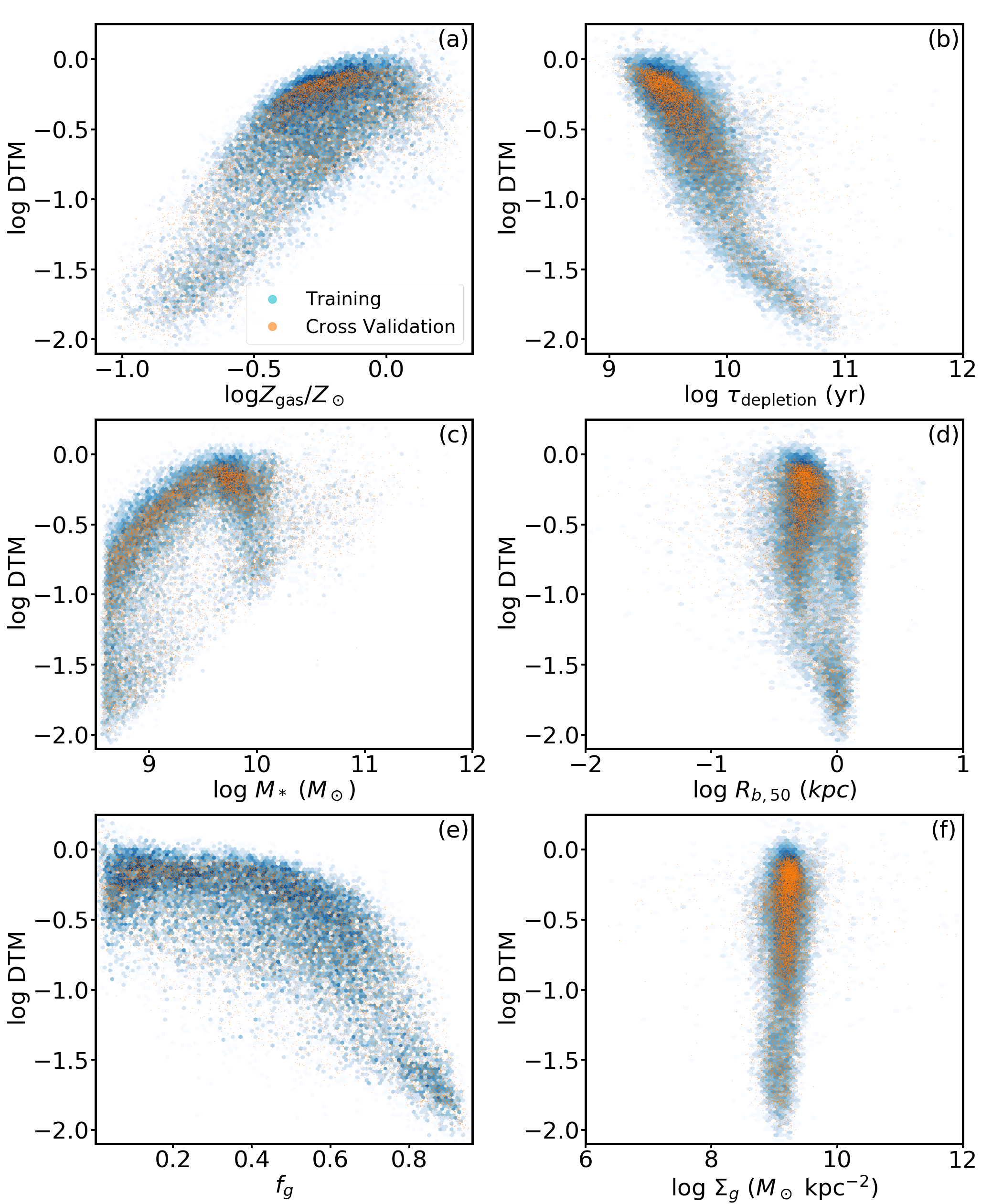}
\caption{ Analogous to Figure~\ref{fig:dtg_fit} but for DTM instead of DGR.
\label{fig:dtm_fit}}
\end{figure}

The ERT is set up using $70\%$ of the selected \simba\ galaxy sample as a training set, with a maximum depth of 20 levels.  We then use the remaining 30\% for validation.  The algorithm then generates a mapping between the inputs and the two desired outputs (DGR and DTM).  By using ERT, we also have access to importance levels, which are determined as the relative depth of a given input parameter was used for branching the tree.

Figure~\ref{fig:dtg_fit} shows the fitted relation between the DGR and various galaxy physical properties.  Cyan points show the training set, and orange points denote the validation set; taken together, they represent all simulation star-forming galaxies.  Observations are shown from \citet{DeVis2019} as black points. 

Comparing to observations, \simba\ reproduces the observed $z=0$ DGR values as a function of various galaxy physical quantities reasonably well.  The DGR increases with metallicity and stellar mass, though less tightly so with the latter.  The DGR also drops with the gas fraction and depletion time, probably reflecting underlying trends from the stellar mass dependence. The DGR shows no clear dependence on $R_{b,50}$ or $\Sigma_g$. For $Z_{\rm gas}$, $\tau_{\rm depletion}$, $M_*$ and $f_g$, the Spearman's rank correlation coefficients are $r=0.87, -0.63, 0.64, 0.81$, respectively, compared to 0.33 for $R_{b,50}$ and 0.14 for $\Sigma_g$.

Figure~\ref{fig:dtm_fit} shows the analogous plot for the DTM ratio.  The trends are broadly similar, with DTM increasing with $Z_{\rm gas}$ and $M_*$, and decreasing with $f_g$ and $\tau_{\rm depletion}$.   This suggests that $M_*$, $f_g$, and $\tau_{\rm depletion}$ may provide additional information that will enable tighter prediction of both the DGR and the DTM.  Rank correlation coefficients are similar to the DGR case.

\begin{figure}
\begin{minipage}{0.52\textwidth}
\includegraphics[width =  0.95 \textwidth]{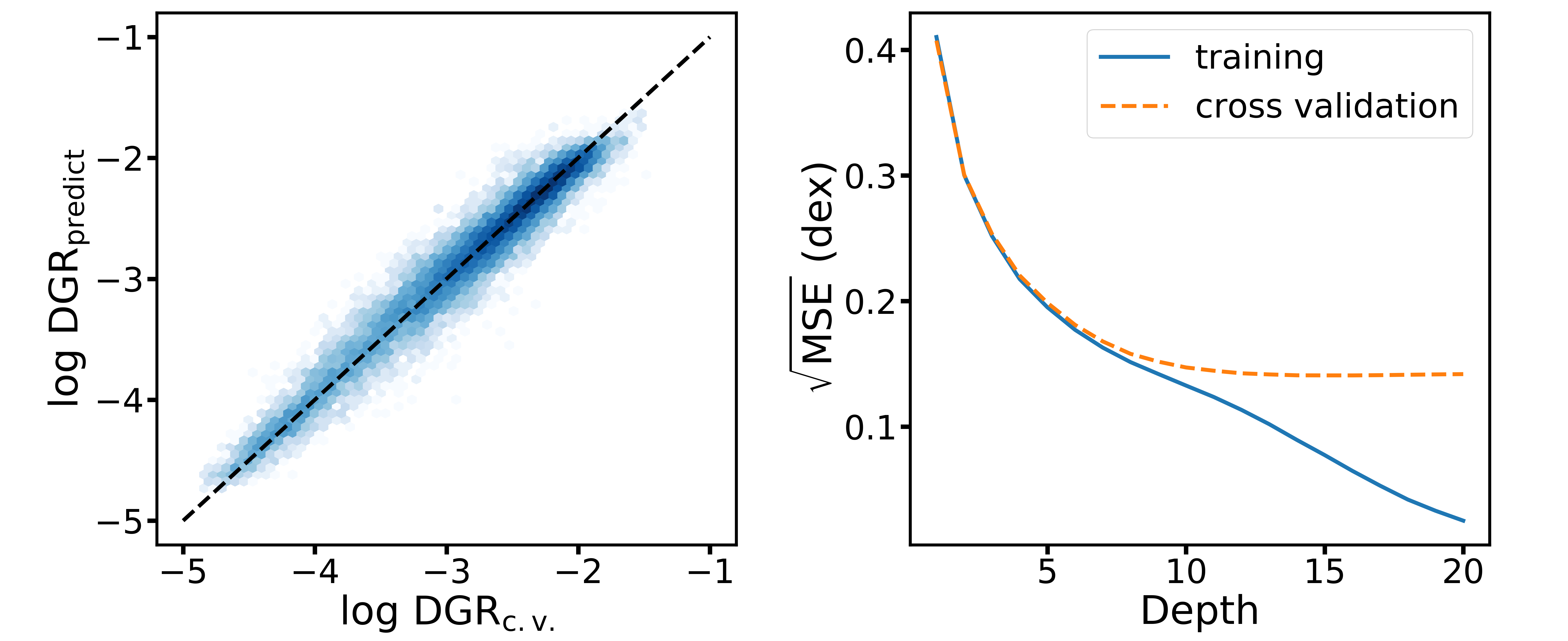}
\end{minipage}
\begin{minipage}{0.52\textwidth}
\includegraphics[width = 0.95 \textwidth]{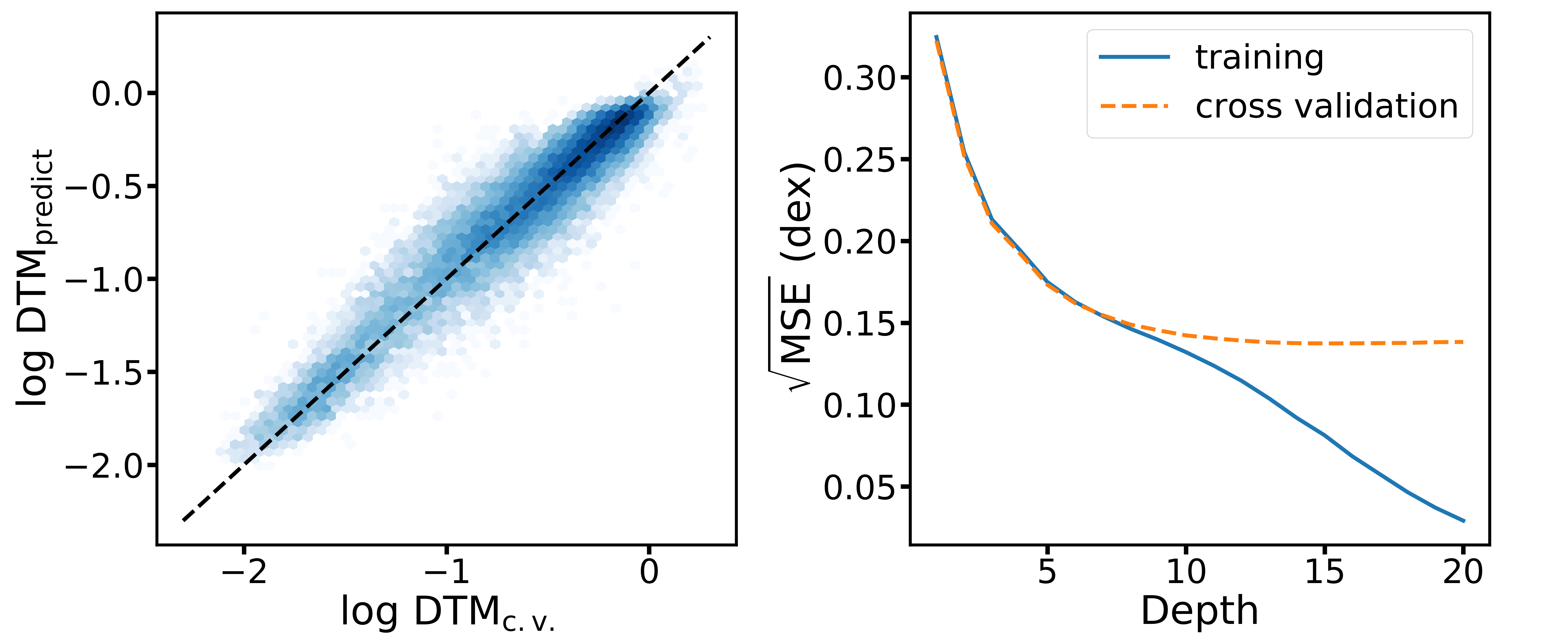}
\end{minipage}

\caption{Top left: a hexbin plot of the predicted DGR derived from physical properties of galaxies in the cross-validation set ${\rm DGR_{predict}}$ and their ``real'' ${\rm DGR_{c.v.}}$ from the simulation at $z=0$. Top right: mean squared error (MSE) of the predicted DGR compared to the ``real'' DGR as a function of maximum depths of ERT for both the training set (blue line) and the cross-validation set (orange dashed line). Bottom panels: analogous to the top panels but for DTM. We use an MSE of the training set to  measure the bias of the model (i.e. to what degree can the model fit the real data) and MSE of the cross-validation set to measure the variance (i.e. how sensitive the model is to noises). We choose the optimal depth 20 by trading off biases and variances
\label{fig:dtg_cv}}
\end{figure}

Since we know the true values for the 30\% validation set, we can examine how well the ERT is able to reproduce these true values.  The quality of fitting is shown in the left panels of Figure~\ref{fig:dtg_cv}, for the DGR (top panels) and DTM (bottom). We find a very tight relation with a $\sim 0.15$~dex scatter estimated by the mean squared error (MSE) between the predicted DGR and true (simulated) DGR of the cross-validation set. This scatter is significantly reduced from $\sim 0.28$~dex when only $Z_{\rm gas}$ is used, showing that the machine learning is effective at generating better predictions for the DGR.  Similarly, for the DTM, the scatter is reduced from 0.27~dex when only $Z_{\rm gas}$ is used, to $0.14$ using the full ERT mapping.

To examine the sensitivity to the ERT tree depth, we show in the right panel of Figure~\ref{fig:dtg_cv} the MSE as a function of tree depth.  We see that increasing the tree depth initially greatly improves predictions, but beyond a depth of $\ga 9$ levels, there is essentially no improvement. This is true for both the DGR and the DTM ratio. At this point, given the sample size and the number of parameters, there is no more information contained in additional tree levels.  Hence we find an optimal maximum ERT tree depth of 9 levels for this sample.

\begin{figure}
\begin{minipage}{0.23\textwidth}
\includegraphics[width = 1.05 \textwidth]{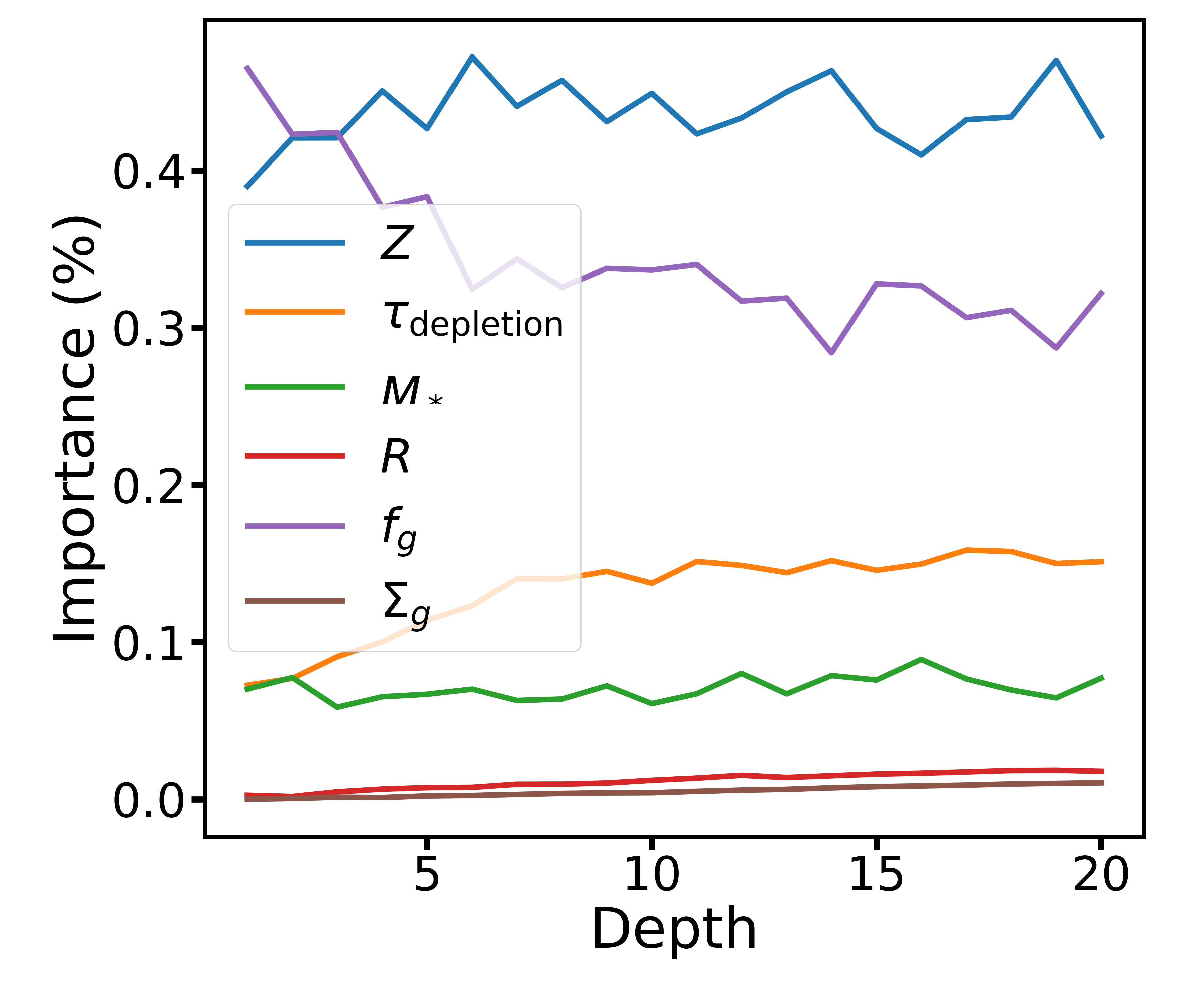}
\end{minipage}
\begin{minipage}{0.23\textwidth}
\includegraphics[width = 1.05 \textwidth]{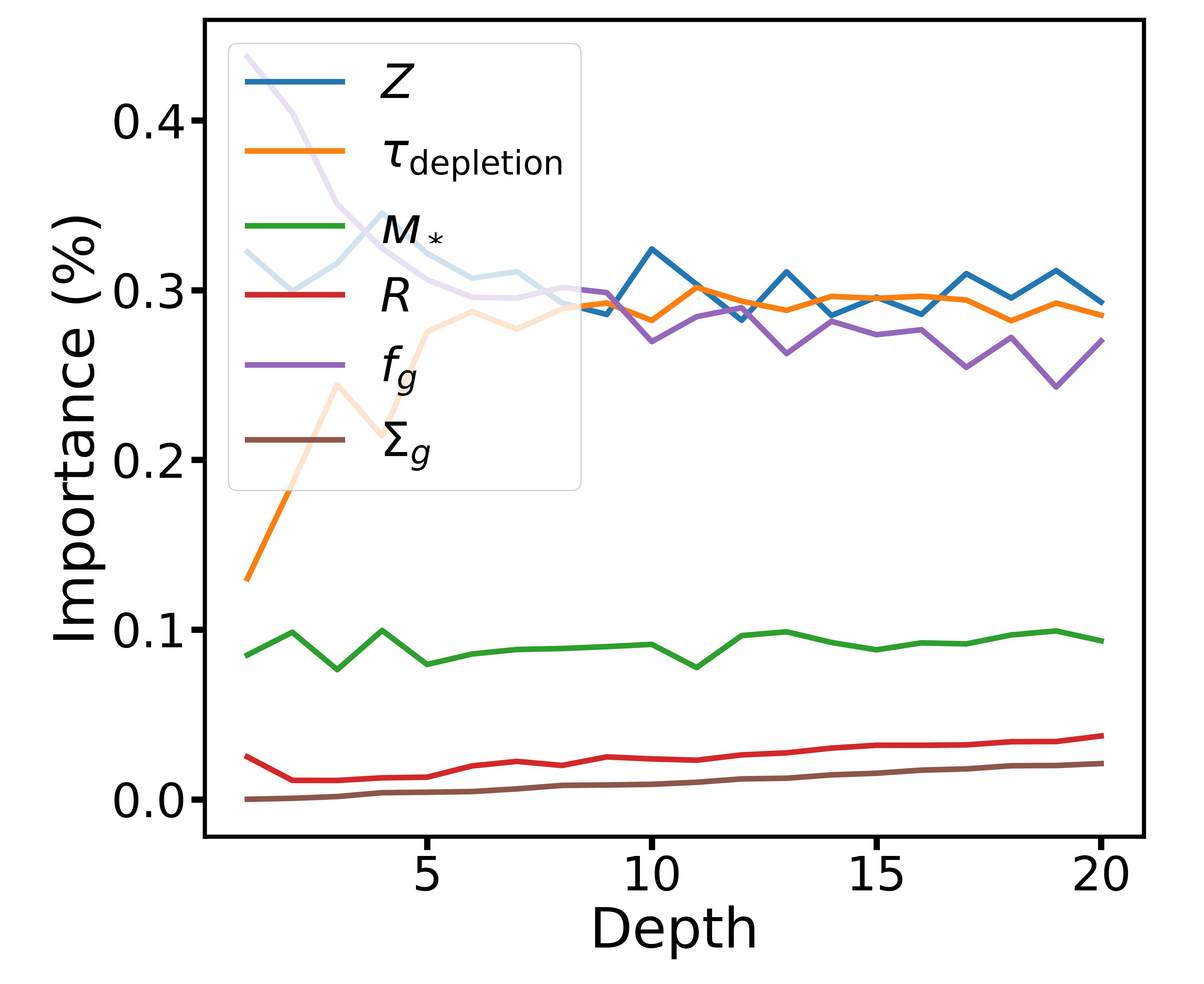}
\end{minipage}
\caption{Left: the relative importance of different galaxy properties in predicting DGR. Right: the relative importance of different galaxy properties in predicting DTM. For both DGR and DTM, four most important properties are $Z_{\rm gas}$, $\tau_{\rm depletion}$, $f_g$ and $M_*$.
\label{fig:dtg_imp}}
\end{figure}

Finally, we examine importance levels of the input physical parameters, as returned by the ERT algorithm, shown in Figure~\ref{fig:dtg_imp}.  At the optimal depth of 9 levels, the left panel shows that the DGR appears to be most directly correlated with the metallicity, followed by the gas fraction. The depletion time and stellar mass also assist with the fitting at a lower level.  As expected, the half-mass radius and the gas surface density do not contribute significant information.

The trends are broadly similar for the DTM, as shown in the right panel of Figure~\ref{fig:dtg_imp}.  However, it now appears that the metallicity, depletion time, and gas fractions all show similar levels of importance.  The stellar mass still shows lower importance, and $R_{b,50}$ and $\Sigma_g$ are again irrelevant.  

In summary, our ERT-based machine learning framework is able to significantly improve the predictive power for the DGR and DTM relations.  Using only $Z_{\rm gas}$ results in scatters of $\sim 0.3$~dex, while using the ERT-generated mapping reduces the scatter to $\sim 0.15$~dex. The key quantities driving this are the metallicity, gas fraction, depletion time, and (to a lesser extent) stellar mass, while in \simba\ the dependencies on the baryonic half-mass radius and gas surface density are negligible.  The map determined via ERT using \simba\ can be applied by modelers who usually have no information about dust or do not track dust evolution in a self-consistent way, which will provide a more accurate estimation of dust mass than that based on a simple assumption of DGR (or DTM).  Alternatively, observers with information about these global galaxy quantities can utilise these algorithms to estimate the dust-to-gas or dust-to-metals ratios in their galaxies.  

\section{Discussion}
\label{sec:discussion}

\subsection{Physical Underpinnings}

We now take a deeper dive in to the details of the trends of the DGR with various physical properties, by examining the correlations of the DGR versus various galaxy physical quantities as shown in Figure~\ref{fig:dtg_fit}.  We focus our discussion on the DGR, as the trends and interpretations for the DTM ratio are analogous.

In panel (a) of Figure~\ref{fig:dtg_fit} we show the DGR-$Z_{\rm gas}$ relation. At $Z < 0.2 Z_\odot$, the relation is roughly linear, corresponding to an approximately constant but low DTM (c.f Figure~\ref{fig:dtm_fit}a). While there is a steep increase from  $Z\sim 0.2 Z_\odot$ to $Z \sim 0.5 Z_\odot$, the relation is again roughly linear at $Z > 0.5 Z_\odot$, corresponding to an approximately constant DTM $\sim 0.8$ (see Figure~\ref{fig:dtm_fit}a). These three different trends shows three different regimes of dust enrichment. At lower $Z_{\rm gas}$, the galaxies are under-evolved and the dust enrichment is dominated by dust production via condensation of ejecta from late-stage stars. In the intermediate $Z_{\rm gas}$ regime, dust growth via accreting gas-phase metals gradually take over the enrichment process. At higher $Z_{\rm gas}$, the growth is extremely strong and the dust mass is mainly determined by the gas-phase metals available for accretion.

In panel (c) of Figure~\ref{fig:dtg_fit}, we show the DGR-$M_*$ relation. The relation flattens after $M_*\ga 10^{9.5} M_\odot$. This transition follows the flattening of the metal-metallicity relations above a comparable $M_*$ (\citealt{Tremonti2004}; see Figure~9 in \citealt{Dave2019}) plus the fact that the DGR primarily traces the galaxy metallicity as Figure~\ref{fig:dtg_fit}(a) shows.  At higher masses, most galaxies are quenched, so the DGR actually drops and has a large scatter owing to dust destruction.

In Figure~\ref{fig:dtg_fit}(b) and Figure~\ref{fig:dtg_fit}(e) we show the DGR-$\tau_{\rm depletion}$ and DGR-$f_g$ relation respectively. Galaxies with lower $f_g$ and $\tau_{\rm depletion}$, which implies that they are quiescent and highly evolved, tend to have a higher DGR. The relations are flatter at low $f_g$ and $\tau_{\rm depletion}$, because the rapid grain growth due to abundant metals is countered by enhanced destructive processes, i.e. shock waves from supernovae, thermal sputtering and astration. These trends shows the correlation between DGR and galaxy evolutionary stages which proceeds as star formation deplete gas and build up metals.  We note, that there are high-$M_*$ (usually high-$Z_{\rm gas}$) galaxies, whose $f_g$ is still relatively high and whose star formation rates are not highly suppressed, having a relatively low DGR. A similar situation applies to some low-$M_*$ galaxies. This contributes to the scatter in the DGR-$Z_{\rm gas}$ plane.

We conclude that the DGR (and DTM) can be determined by $Z_{\rm gas}$ along with $M*$ which reflects the chemical enrichment history and $\tau_{\rm SF}$ along with $f_g$ which indicates the evolutionary stage of galaxies. 
Physically, this suggests that metal enrichment history reflected by $Z_{\rm gas}$ and $M_*$, and evolutionary stages quantified by $\tau_{\rm depletion}$ (see also \citealt{Asano2013,Zhukovska2014,Feldmann2015}) and $f_g$ (see also \citealt{DeVis2019}) are the main drivers of the scatter in the DGR$-Z_{\rm gas}$ plane.  Meanwhile, galaxy compactness as quantified by $R_{b,50}$ and $\Sigma_g$ does not seem to impact the DGR or DTM, showing that at least in \simba\ dust content is insensitive to galactic structure -- with the caveat that given \simba's $\sim1$~kpc resolution, galactic structure may not be faithfully modeled in detail. 

\subsection{Comparison with other models}
\label{sec:compare}
The DMF and the DGR-$Z_{\rm gas}$ relation have been studied by cosmological hydrodynamic 
simulations \citep{McKinnon2017,Hou2019} and semi-analytic models 
\citep{Popping2017}. Like \citet{McKinnon2017} and 
\citet{Popping2017}, our work predicts that the dust mass function increases monotonically from z = 2 to 0 at the high-mass end ($M_d \gtrsim 10^{8} M_\odot$), and is unable to simultaneously match the  $z=0$ to $z=2$ DMF. Nevertheless, our result appears to have the closest match to observations to date, underpredicting the $z=2$ DMF by a factor $\sim 3$, which is significantly better than \citet{McKinnon2017} where galaxies with $M_d \gtrsim 10^{8} M_\odot$ are hardly produced. On the other hand, \citet{Hou2019} tracks two types of grains (small/large) and grain-grain shattering and coagulation. They also implement a subgrid-model to boost the density of unresolved dense gas, which is not adopted by our model (therefore we use a short $\tau_{\rm ref}$) and a simple AGN feedback model to suppress star forming activity of massive galaxies. They are able to reproduce the non-monotonic trend of the high-mass-end DMF evolution from $z=2$ to $z=0$, which was observed by \citet{Dunne2011}. However, they fail to match the observed DMF at $z=0$ (where they overproduce high $M_d$ galaxies) to $z=2$ (where they underproduce high $M_d$ galaxies).

What steps forward are necessary for simulations to match the observed $z=2$ DMF? Some possible solutions include: (1) implementing dust yields in stellar ejecta as a function of a star's mass and metallicity \citep{Ferrarotti2006,Bianchi2007,Zhukovska2007,Nanni2013,Schneider2014}, or the local ISM density or temperature, as pointed out by \citet{McKinnon2017}; 
(2) tracking the evolution of grains with different sizes (see e.g. \citealt{Hou2019} for 
two-size grains and \citealt{McKinnon2018,Aoyama2019} for a continuous distribution of grain sizes). 
Solution \#2 will be implemented in an upcoming paper. We expect that a non-monotonic 
evolution of high-mass-end DMF would result from the intensified grain-grain collisions at lower redshift which will
generally lead to an increasing abundance of smaller grains that experience faster destruction.

\begin{figure}
\includegraphics[width=0.5 \textwidth]{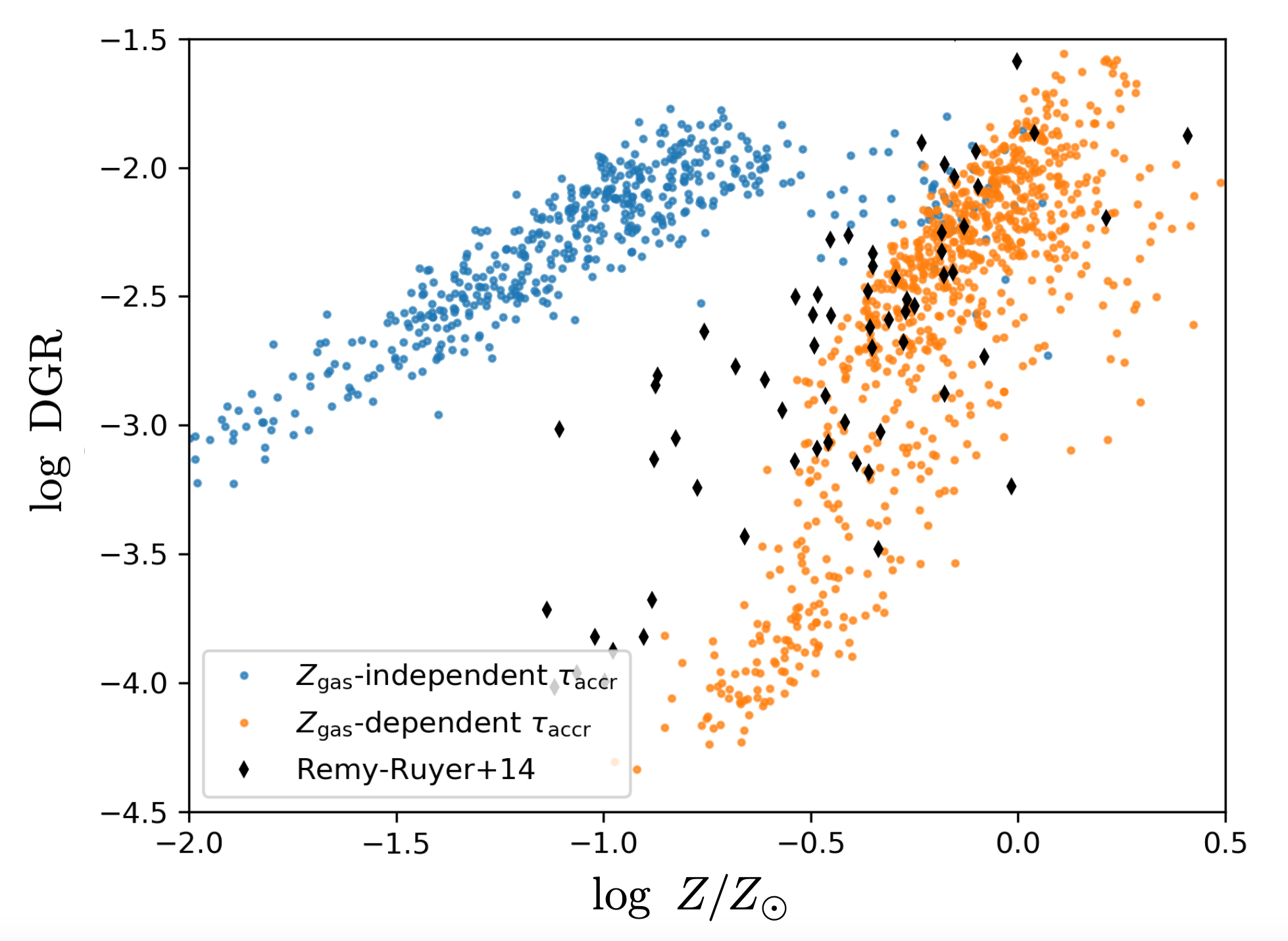}
\caption{A comparison of DGR-$Z_{\rm gas}$ relations from two test runs where accretion timescales $\tau_{\rm accr}$ (c.f. Equation~\ref{eq:tauaccr}) are dependent (as adopted by \citealt{Popping2017}, \citealt{Hou2019} and our work) or independent (as adopted by \citealt{McKinnon2017}) of $Z_{\rm gas}$. These test runs have the same mass resolution as the primary \simba\ run, but have $256^3$ dark matter particles and $256^3$ gas elements in a cube of $25\hmpc$. It shows $Z_{\rm gas}$-dependent $\tau_{\rm accr}$ is essential to reproduce the observed relation.
\label{fig:dis_dtg}}
\end{figure}

Turning to the DGR-$Z_{\rm gas}$ relation: The simulations by \citet{McKinnon2017} obtain a rather flat relation between the DGR and gas-phase metallicity from low $Z_{\rm gas}$ to high $Z_{\rm gas}$, mainly because their accretion timescale for dust growth (see Equation (5) of \citealt{McKinnon2016}) does not vary with the 
local ISM $Z_{\rm gas}$.  We find that the dependency of the accretion timescale on metallicity is essential to reproduce the observed DGR-$Z_{\rm gas}$ relation as is shown in \citet{Popping2017}, 
\citet{Hou2019} and our work, in our case by the results of two test runs in Figure~\ref{fig:dis_dtg}.  If we assume a metallicity-independent accretion timescale, then we get a flatter and higher relation (blue points) compared to the \citet{Remy-Ruyer2014} observations.  The actual observations are slightly shallower than the predictions from the metallicity-dependent model, suggesting that perhaps our metallicity dependence should be softened somewhat.

Similar to \citet{Popping2017}, our results show a weak evolution of the DGR-$Z_{\rm gas}$ relation from $z=0$ to $z=6$, especially for high metallicity galaxies.  This is encouraging given that the underlying galaxy formation models (SAM vs hydrodynamic) are rather different from one another.  Beyond this, the treatment of dust growth in both models is different: grain growth by accretion in our work does not use the information of any detailed subgrid ISM model, whereas in the \citet{Popping2017}  SAM, the accretion time-scale is calculated by inferring the gas density in molecular clouds from SFR laws.  The fact that these two very different models arrive at the same conclusion for a relatively modest evolution in the DGR-$Z$ relation likely underscores its robustness.   On the other hand, \citet{Hou2019} shows that the DGR at a fixed $Z_{\rm gas}$ builds up significantly a factor of $\sim5$ from $z=5$ to $z=0$.  This discrepancy mainly comes from the different treatment of grain growth and feedback. Grain growth in our simulation is overall stronger, therefore our simulated galaxies are able to reach the quasi-static states within a shorter period of time. Moreover, \simba\ has a more sophisticated feedback mechanisms that suppress further metal enrichment at lower redshifts, particularly in massive galaxies.

\citet{Hou2019} explores the scaling relations with a variety of galaxy properties other than $Z_{\rm gas}$.  Comparing Figure~\ref{fig:dtg_fit} (c) and (e) against their Figure~4 (b) and (d), we find that the trend predicted by both simulations are similar but there are big discrepancies at the high $M_*$ ($M_* > 10^{9.5} M_\odot$, the turning point of mass-metallicity relation) and low $f_g$ end, where \citet{Hou2019}'s model overproduces the DGR compared against observational constraints. In this regime, our DGR-$M_*$ and DGR-$f_g$ relations flatten. The sophisticated black hole feedback model in \simba\, compared to the simple phenomenological AGN feedback model used in their work which may underestimate the suppressing power, can explain the difference. A similar analysis applies to the low $f_g$ regime where most of the galaxies are highly evolved and massive, and black hole feedback mechanisms are influential.

\subsection{Caveats}
Here we point out some caveats of our simulation. First, we note that the choice of free parameters for dust production, grain growth and grain destruction via SNe shocks are not well constrained (c.f. Table~\ref{tab:1}) and most likely degenerate. Though grain growth dominates the evolution of dust content, a combination of free parameters different from what is chosen in this work could potentially lead to an equally good (or even slightly better) match to observations, e.g. one with stronger production, weaker grain growth and stronger destruction, or one with stronger grain growth and much stronger destruction. On the other hand, the observed metallicities depend strongly on which strong-line calibration is used. The match to \cite{DeVis2019} would be worse if e.g. calibrations of \cite{Pettini2004} are used instead of \cite{Pilyugin2016}.

Besides, dust plays an important role in cooling and shielding gas, catalyzing important chemical reactions (e.g. formation of molecular hydrogen) and recombination processes, and alternating the interstellar radiation fields. However, dust in our simulation only affects the ISM by depleting gas-phase metals thus reducing the efficiency of gas cooling channels.  Work to implement the dust physics in a fully self-consistent manner is under way.

Beyond this, in our current model implementation, dust grains are fully coupled to gas flows. In reality, dust grains can be decoupled from gas due to radiative forces and lack of pressure and experience/apply drag forces from/to gas \citep{Squire2018,Hopkins2018a,Hopkins2018b}. Dust grains in this work are assumed to have the same grain size to capture the major processes that evolve the dust mass. In the future we will implement ``active'' dust particles that are not strictly coupled to gas to track the evolution of grain size distributions and take grain-grain collisions into account. As mentioned in \S\ref{sec:compare}, we expect the evolution of the grain size distribution would alter the evolution of the cosmic dust content. The use of dust super-particles to sample the spatial distribution of dust would also save the memory usage, making it computationally feasible to track multiple-size grains in a large volume cosmological simulation, even though we expect the effect on dynamics would be negligible because of the lack of spatial resolution of our large-volume cosmological simulations and strong radiative fields.

Finally, the 100$\hmpc$ \simba\ volume lacks the resolution to resolve a multi-phase ISM, which is a common issue for large-volume cosmological simulations.  As a result, parameters such as the reference accretion time scale $\tau_{\rm ref}$ have to be tuned such that the effective gas density is boosted. We also assume fixed dust destruction and condensation efficiencies, which may  actually be functions of local ISM properties \citep{Seab1983,McKee1987,Ferrarotti2006,Bianchi2007,Zhukovska2007,Yamasawa2011,Nanni2013,Schneider2014,Temim2015}. Calibrating parameters we have used against numerical simulations of resolved ISM is one potential approach to make improvement.

\section{Summary}
\label{sec:conclude}

We have developed a self-consistent model for the formation, growth and destruction of dust in the \simba\ cosmological galaxy formation simulation, and used these to compare to predictions to observed predictions, as well as 
study the physical drivers of the dust-to-gas and dust-to-metals ratios in galaxies.  We also develop a machine learning framework to relate the dust-to-gas ratio (DGR) and dust-to-metal ratio (DTM) to various input global galaxy properties. Our main results are as follows:

Our main results follow:
\begin{itemize}
\item \simba\ broadly reproduces the observed dust mass functions across the measured redshifts of $z=0-2$, albeit with modest under-prediction at high-$z$.  The low-mass end steepens at high redshift.

\item We find a relationship between the dust-to-gas ratio of star-forming galaxies and their gas phase metallicity such that lower metallicity galaxies have lower dust-to-gas ratios.  This is broadly in accord with observations.  There is little evolution with redshift in this relationship.  Meanwhile, quenched galaxies show lower DGR and essentially no relationship in this space (Figure~\ref{fig:dtgz}).  This non-constant dust-to-metals ratio with metallicity (eq.~\ref{equation:DGR_fit}) has implications for galaxy formation models that historically have assumed a constant dust-to-metals ratio  \citep[see e.g.][]{Silva1998,Granato2000,Baugh2005,Lacey2010,Narayanan2010,Narayanan2015,narayanan18a,narayanan18b,
Fontanot2011,Niemi2012,Somerville2012,Hayward2013,Cowley2017,Katz2019,Ma2019}.
    
\item The DTM ratio vs. metallicity relation drops at low metallicity, akin to the DGR relationship (Figure~\ref{fig:dtmz}).  This is consistent with low redshift result, yet may be in tension with observational constraints at low metallicities at high redshifts from GRBs and DLAs. Other trends are qualitatively similar to those of the DGR.

\item In order to help both modelers and observers estimate more accurate DGR and DTM ratios, we have developed a publicly-available machine learning framework that generates a mapping between the DGR and DTM ratio and a set of galaxy physical properties, showing that it depends significantly on various galaxy properties.  The machine learning framework reduces the scatter in the prediction from $\sim 0.3$~dex in the DTM and DGR using a simple fit to $Z_{\rm gas}$ alone, down to $\sim 0.15$~dex using the machine learning framework.  This code is available at \url{https://bitbucket.org/lq3552/dust_galaxy_analyzer}. 

\item While the DGR and DTM ratios depend most sensitively on the gas phase metallicity in galaxies, we demonstrate that there are important secondary relationships between these ratios and the depletion time scale, stellar mass and gas fraction of galaxies (Figure~\ref{fig:dtg_fit}). The DGR and DTM ratio both drop to lower $M_*$, and rise to lower $f_{\rm gas}$ and gas depletion times.  There is no dependence on gas surface density and baryonic half-mass radius.  Hence dust content is governed by both long-term evolutionary processes such as metal content and stellar mass, as well as short-term variations such as varying gas content and (commensurately) depletion times.

Overall, the \simba\ dust model is at least as successful compared with other current dust models implemented in cosmological simulations.  However, there remain various caveats and potential directions for improvement.  These include having active dust (not tied to the gas), multiple dust grain sizes, and implementing more sophisticated dust cooling.  Furthermore, there are various free parameters that are constrained indirectly by observations, which might be better constrained using high-resolution ISM simulation.  By using such a multi-scale approach to combine high resolution simulations and observational constraints into a cosmological galaxy formation model, we are moving towards more comprehensively studying the evolution of galaxy dust on cosmological scales.

\end{itemize}

\section*{Acknowledgements}

The authours acknowledge helpful discussions with Ryan McKinnon, Gergo Popping,  and Paul Torrey, and thank the anonymous referee for constructive comments.  QL acknowledges support from the University of Florida Informatics Institute.  QL and DN acknowledge support from the NSF via grant AST-1715206, and the Space Telescope Science Institute via HST AR-15043.0001.
RD acknowledges support from the Wolfson Research Merit Award program of the U.K. Royal Society.
This work used the DiRAC@Durham facility managed by the Institute for Computational Cosmology on behalf of the STFC DiRAC HPC Facility. The equipment was funded by BEIS capital funding via STFC capital grants ST/P002293/1, ST/R002371/1 and ST/S002502/1, Durham University and STFC operations grant ST/R000832/1. DiRAC is part of the National e-Infrastructure.  Some of the simulations presented here were performed on the University of Florida HiPerGator2.0 supercomputing facility, and the authours are grateful to the staff that supports the software and hardware on HiPerGator.

\bibliographystyle{mnras}
\bibliography{ms}

\begin{thebibliography}{}
\makeatletter
\relax
\def\mn@urlcharsother{\let\do\@makeother \do\$\do\&\do\#\do\^\do\_\do\%\do\~}
\def\mn@doi{\begingroup\mn@urlcharsother \@ifnextchar [ {\mn@doi@}
  {\mn@doi@[]}}
\def\mn@doi@[#1]#2{\def\@tempa{#1}\ifx\@tempa\@empty \href
  {http://dx.doi.org/#2} {doi:#2}\else \href {http://dx.doi.org/#2} {#1}\fi
  \endgroup}
\def\mn@eprint#1#2{\mn@eprint@#1:#2::\@nil}
\def\mn@eprint@arXiv#1{\href {http://arxiv.org/abs/#1} {{\tt arXiv:#1}}}
\def\mn@eprint@dblp#1{\href {http://dblp.uni-trier.de/rec/bibtex/#1.xml}
  {dblp:#1}}
\def\mn@eprint@#1:#2:#3:#4\@nil{\def\@tempa {#1}\def\@tempb {#2}\def\@tempc
  {#3}\ifx \@tempc \@empty \let \@tempc \@tempb \let \@tempb \@tempa \fi \ifx
  \@tempb \@empty \def\@tempb {arXiv}\fi \@ifundefined
  {mn@eprint@\@tempb}{\@tempb:\@tempc}{\expandafter \expandafter \csname
  mn@eprint@\@tempb\endcsname \expandafter{\@tempc}}}

\bibitem[\protect\citeauthoryear{{Agarwal}, {Dav{\'e}}  \& {Bassett}}{{Agarwal}
  et~al.}{2018}]{Agarwal2018}
{Agarwal} S.,  {Dav{\'e}} R.,   {Bassett} B.~A.,  2018, \mn@doi [\mnras]
  {10.1093/mnras/sty1169}, \href
  {https://ui.adsabs.harvard.edu/abs/2018MNRAS.478.3410A} {478, 3410}

\bibitem[\protect\citeauthoryear{Angl{\'e}s-Alc{\'a}zar, Dav{\'e},
  Faucher-Gigu{\`e}re, {\"O}zel  \& Hopkins}{Angl{\'e}s-Alc{\'a}zar
  et~al.}{2017a}]{Angles2017a}
Angl{\'e}s-Alc{\'a}zar D.,  Dav{\'e} R.,  Faucher-Gigu{\`e}re C.-A.,  {\"O}zel
  F.,   Hopkins P.~F.,  2017a, \mn@doi [\mnras] {10.1093/mnras/stw2565}, 464,
  2840

\bibitem[\protect\citeauthoryear{Angl{\'e}s-Alc{\'a}zar, Faucher-Gigu{\`e}re,
  Kere{\v s}, Hopkins, Quataert  \& Murray}{Angl{\'e}s-Alc{\'a}zar
  et~al.}{2017b}]{Angles2017b}
Angl{\'e}s-Alc{\'a}zar D.,  Faucher-Gigu{\`e}re C.-A.,  Kere{\v s} D.,  Hopkins
  P.~F.,  Quataert E.,   Murray N.,  2017b, \mn@doi [\mnras]
  {10.1093/mnras/stx1517}, 470, 4698

\bibitem[\protect\citeauthoryear{Aoyama, Hou, Shimizu, Hirashita, Todoroki,
  Choi  \& Nagamine}{Aoyama et~al.}{2017}]{Aoyama2017}
Aoyama S.,  Hou K.-C.,  Shimizu I.,  Hirashita H.,  Todoroki K.,  Choi J.-H.,
  Nagamine K.,  2017, \mn@doi [\mnras] {10.1093/mnras/stw3061}, 466, 105

\bibitem[\protect\citeauthoryear{Aoyama, Hou, Hirashita, Nagamine  \&
  Shimizu}{Aoyama et~al.}{2018}]{Aoyama2018}
Aoyama S.,  Hou K.-C.,  Hirashita H.,  Nagamine K.,   Shimizu I.,  2018,
  \mn@doi [\mnras] {10.1093/mnras/sty1431}, 478, 4905

\bibitem[\protect\citeauthoryear{{Aoyama}, {Hirashita}  \& {Nagamine}}{{Aoyama}
  et~al.}{2019}]{Aoyama2019}
{Aoyama} S.,  {Hirashita} H.,   {Nagamine} K.,  2019, arXiv e-prints, \href
  {https://ui.adsabs.harvard.edu/abs/2019arXiv190601917A} {p. arXiv:1906.01917}

\bibitem[\protect\citeauthoryear{{Asano}, {Takeuchi}, {Hirashita}  \&
  {Inoue}}{{Asano} et~al.}{2013}]{Asano2013}
{Asano} R.~S.,  {Takeuchi} T.~T.,  {Hirashita} H.,   {Inoue} A.~K.,  2013,
  \mn@doi [Earth, Planets, and Space] {10.5047/eps.2012.04.014}, \href
  {https://ui.adsabs.harvard.edu/\#abs/2013EP&S...65..213A} {65, 213}

\bibitem[\protect\citeauthoryear{{Asplund}, {Grevesse}, {Sauval}  \&
  {Scott}}{{Asplund} et~al.}{2009}]{Asplund2009}
{Asplund} M.,  {Grevesse} N.,  {Sauval} A.~J.,   {Scott} P.,  2009, \mn@doi
  [\araa] {10.1146/annurev.astro.46.060407.145222}, \href
  {https://ui.adsabs.harvard.edu/abs/2009ARA&A..47..481A} {47, 481}

\bibitem[\protect\citeauthoryear{Ball, Brunner, Myers, Strand, Alberts, Tcheng
  \& Llor{\`a}}{Ball et~al.}{2007}]{Ball2007}
Ball N.~M.,  Brunner R.~J.,  Myers A.~D.,  Strand N.~E.,  Alberts S.~L.,
  Tcheng D.,   Llor{\`a} X.,  2007, \mn@doi [\apj] {10.1086/518362}, 663, 774

\bibitem[\protect\citeauthoryear{{Barlow}}{{Barlow}}{1978}]{Barlow1978}
{Barlow} M.~J.,  1978, \mn@doi [\mnras] {10.1093/mnras/183.3.367}, \href
  {https://ui.adsabs.harvard.edu/\#abs/1978MNRAS.183..367B} {183, 367}

\bibitem[\protect\citeauthoryear{Baugh, Lacey, Frenk, Granato, Silva, Bressan,
  Benson  \& Cole}{Baugh et~al.}{2005}]{Baugh2005}
Baugh C.~M.,  Lacey C.~G.,  Frenk C.~S.,  Granato G.~L.,  Silva L.,  Bressan
  A.,  Benson A.~J.,   Cole S.,  2005, \mn@doi [\mnras]
  {10.1111/j.1365-2966.2004.08553.x}, 356, 1191

\bibitem[\protect\citeauthoryear{{Beeston} et~al.,}{{Beeston}
  et~al.}{2018}]{Beeston2018}
{Beeston} R.~A.,  et~al., 2018, \mn@doi [\mnras] {10.1093/mnras/sty1460}, \href
  {https://ui.adsabs.harvard.edu/abs/2018MNRAS.479.1077B} {479, 1077}

\bibitem[\protect\citeauthoryear{Bekki}{Bekki}{2015}]{Bekki2015}
Bekki K.,  2015, \mn@doi [\mnras] {10.1093/mnras/stv165}, 449, 1625

\bibitem[\protect\citeauthoryear{{Bendo} et~al.,}{{Bendo}
  et~al.}{2010}]{Bendo2010}
{Bendo} G.~J.,  et~al., 2010, \mn@doi [\mnras]
  {10.1111/j.1365-2966.2009.16043.x}, \href
  {http://adsabs.harvard.edu/abs/2010MNRAS.402.1409B} {402, 1409}

\bibitem[\protect\citeauthoryear{{Berry}, {Somerville}, {Haas}, {Gawiser},
  {Maller}, {Popping}  \& {Trager}}{{Berry} et~al.}{2014}]{Berry2014}
{Berry} M.,  {Somerville} R.~S.,  {Haas} M.~R.,  {Gawiser} E.,  {Maller} A.,
  {Popping} G.,   {Trager} S.~C.,  2014, \mn@doi [\mnras]
  {10.1093/mnras/stu613}, \href
  {https://ui.adsabs.harvard.edu/abs/2014MNRAS.441..939B} {441, 939}

\bibitem[\protect\citeauthoryear{Bianchi \& Ferrara}{Bianchi \&
  Ferrara}{2005}]{Bianchi2005}
Bianchi S.,  Ferrara A.,  2005, \mn@doi [\mnras]
  {10.1111/j.1365-2966.2005.08762.x}, 358, 379

\bibitem[\protect\citeauthoryear{{Bianchi} \& {Schneider}}{{Bianchi} \&
  {Schneider}}{2007}]{Bianchi2007}
{Bianchi} S.,  {Schneider} R.,  2007, \mn@doi [\mnras]
  {10.1111/j.1365-2966.2007.11829.x}, \href
  {https://ui.adsabs.harvard.edu/\#abs/2007MNRAS.378..973B} {378, 973}

\bibitem[\protect\citeauthoryear{Bouch{\'e}, Lehnert, Aguirre, P{\'e}roux  \&
  Bergeron}{Bouch{\'e} et~al.}{2007}]{Bouche2007}
Bouch{\'e} N.,  Lehnert M.~D.,  Aguirre A.,  P{\'e}roux C.,   Bergeron J.,
  2007, \mn@doi [\mnras] {10.1111/j.1365-2966.2007.11740.x}, 378, 525

\bibitem[\protect\citeauthoryear{Calura, Pipino  \& Matteucci}{Calura
  et~al.}{2007}]{Calura2008}
Calura F.,  Pipino A.,   Matteucci F.,  2007, \mn@doi [Astronomy {\&}
  Astrophysics] {10.1051/0004-6361:20078090}, 479, 669

\bibitem[\protect\citeauthoryear{Calura, Gilli, Vignali, Pozzi, Pipino  \&
  Matteucci}{Calura et~al.}{2014}]{Calura2014}
Calura F.,  Gilli R.,  Vignali C.,  Pozzi F.,  Pipino A.,   Matteucci F.,
  2014, \mn@doi [\mnras] {10.1093/mnras/stt2329}, 438, 2765

\bibitem[\protect\citeauthoryear{Carrasco~Kind \& Brunner}{Carrasco~Kind \&
  Brunner}{2013}]{Kind2013}
Carrasco~Kind M.,  Brunner R.~J.,  2013, \mn@doi [\mnras]
  {10.1093/mnras/stt574}, 432, 1483

\bibitem[\protect\citeauthoryear{Chiang, Sandstrom, Chastenet, Johnson, Leroy
  \& Utomo}{Chiang et~al.}{2018}]{Chiang2018}
Chiang I.-D.,  Sandstrom K.~M.,  Chastenet J.,  Johnson L.~C.,  Leroy A.~K.,
  Utomo D.,  2018, \mn@doi [\apj] {10.3847/1538-4357/aadc5f}, 865, 117

\bibitem[\protect\citeauthoryear{Clemens et~al.,}{Clemens
  et~al.}{2013}]{Clemens2013}
Clemens M.~S.,  et~al., 2013, \mn@doi [\mnras] {10.1093/mnras/stt760}, 433, 695

\bibitem[\protect\citeauthoryear{Cowley, B{\'e}thermin, Lagos, Lacey, Baugh  \&
  Cole}{Cowley et~al.}{2017}]{Cowley2017}
Cowley W.~I.,  B{\'e}thermin M.,  Lagos C. d.~P.,  Lacey C.~G.,  Baugh C.~M.,
  Cole S.,  2017, \mn@doi [\mnras] {10.1093/mnras/stx165}, 467, 1231

\bibitem[\protect\citeauthoryear{Dav{\'e}, Thompson  \& Hopkins}{Dav{\'e}
  et~al.}{2016}]{Dave2016}
Dav{\'e} R.,  Thompson R.,   Hopkins P.~F.,  2016, \mn@doi [\mnras]
  {10.1093/mnras/stw1862}, 462, 3265

\bibitem[\protect\citeauthoryear{{Dav{\'e}}, {Angl{\'e}s-Alc{\'a}zar},
  {Narayanan}, {Li}, {Rafieferantsoa}  \& {Appleby}}{{Dav{\'e}}
  et~al.}{2019}]{Dave2019}
{Dav{\'e}} R.,  {Angl{\'e}s-Alc{\'a}zar} D.,  {Narayanan} D.,  {Li} Q.,
  {Rafieferantsoa} M.~H.,   {Appleby} S.,  2019, \mn@doi [\mnras]
  {10.1093/mnras/stz937}, \href
  {https://ui.adsabs.harvard.edu/abs/2019MNRAS.486.2827D} {486, 2827}

\bibitem[\protect\citeauthoryear{{De Cia}, Ledoux, Savaglio, Schady  \&
  Vreeswijk}{{De Cia} et~al.}{2013}]{DeCia2013}
{De Cia} A.,  Ledoux C.,  Savaglio S.,  Schady P.,   Vreeswijk P.~M.,  2013,
  \mn@doi [Astronomy {\&} Astrophysics] {10.1051/0004-6361/201321834}, 560, A88

\bibitem[\protect\citeauthoryear{{De Cia}, Ledoux, Mattsson, Petitjean,
  Srianand, Gavignaud  \& Jenkins}{{De Cia} et~al.}{2016}]{DeCia2016}
{De Cia} A.,  Ledoux C.,  Mattsson L.,  Petitjean P.,  Srianand R.,  Gavignaud
  I.,   Jenkins E.~B.,  2016, \mn@doi [Astronomy {\&} Astrophysics]
  {10.1051/0004-6361/201527895}, 596, A97

\bibitem[\protect\citeauthoryear{{De Vis} et~al.,}{{De Vis}
  et~al.}{2017}]{DeVis2017}
{De Vis} P.,  et~al., 2017, \mn@doi [\mnras] {10.1093/mnras/stx981}, \href
  {https://ui.adsabs.harvard.edu/abs/2017MNRAS.471.1743D} {471, 1743}

\bibitem[\protect\citeauthoryear{{De Vis} et~al.,}{{De Vis}
  et~al.}{2019}]{DeVis2019}
{De Vis} P.,  et~al., 2019, \mn@doi [Astronomy {\&} Astrophysics]
  {10.1051/0004-6361/201834444}, 623, A5

\bibitem[\protect\citeauthoryear{Dominik \& Tielens}{Dominik \&
  Tielens}{1997}]{Dominik1997}
Dominik C.,  Tielens A. G. G.~M.,  1997, \mn@doi [\apj] {10.1086/303996}, 480,
  647

\bibitem[\protect\citeauthoryear{Draine}{Draine}{2003}]{Draine2003}
Draine B.~T.,  2003, \mn@doi [\apj] {10.1086/379118}, 598, 1017

\bibitem[\protect\citeauthoryear{Draine \& Salpeter}{Draine \&
  Salpeter}{1979a}]{Draine1979a}
Draine B.~T.,  Salpeter E.~E.,  1979a, \mn@doi [\apj] {10.1086/157165}, 231, 77

\bibitem[\protect\citeauthoryear{{Draine} \& {Salpeter}}{{Draine} \&
  {Salpeter}}{1979b}]{Draine1979b}
{Draine} B.~T.,  {Salpeter} E.~E.,  1979b, \mn@doi [\apj] {10.1086/157206},
  \href {https://ui.adsabs.harvard.edu/\#abs/1979ApJ...231..438D} {231, 438}

\bibitem[\protect\citeauthoryear{Draine et~al.,}{Draine
  et~al.}{2007}]{Draine2007}
Draine B.~T.,  et~al., 2007, \mn@doi [\apj] {10.1086/518306}, 663, 866

\bibitem[\protect\citeauthoryear{Dunne, Eales  \& Edmunds}{Dunne
  et~al.}{2003}]{Dunne2003}
Dunne L.,  Eales S.~A.,   Edmunds M.~G.,  2003, \mn@doi [\mnras]
  {10.1046/j.1365-8711.2003.06440.x}, 341, 589

\bibitem[\protect\citeauthoryear{Dunne et~al.,}{Dunne et~al.}{2011}]{Dunne2011}
Dunne L.,  et~al., 2011, \mn@doi [\mnras] {10.1111/j.1365-2966.2011.19363.x},
  417, 1510

\bibitem[\protect\citeauthoryear{{Dwek}}{{Dwek}}{1998}]{Dwek1998}
{Dwek} E.,  1998, \mn@doi [\apj] {10.1086/305829}, \href
  {https://ui.adsabs.harvard.edu/\#abs/1998ApJ...501..643D} {501, 643}

\bibitem[\protect\citeauthoryear{{Dwek}}{{Dwek}}{2016}]{Dwek2016}
{Dwek} E.,  2016, \mn@doi [\apj] {10.3847/0004-637X/825/2/136}, \href
  {https://ui.adsabs.harvard.edu/\#abs/2016ApJ...825..136D} {825, 136}

\bibitem[\protect\citeauthoryear{{Dwek} \& {Scalo}}{{Dwek} \&
  {Scalo}}{1980}]{Dwek1980}
{Dwek} E.,  {Scalo} J.~M.,  1980, \mn@doi [\apj] {10.1086/158100}, \href
  {https://ui.adsabs.harvard.edu/\#abs/1980ApJ...239..193D} {239, 193}

\bibitem[\protect\citeauthoryear{Eales et~al.,}{Eales et~al.}{2009}]{Eales2009}
Eales S.,  et~al., 2009, \mn@doi [\apj] {10.1088/0004-637X/707/2/1779}, 707,
  1779

\bibitem[\protect\citeauthoryear{Ester, Kriegel, Sander  \& Xu}{Ester
  et~al.}{1996}]{Ester1996}
Ester M.,  Kriegel H.~P.,  Sander J.,   Xu X.,  1996, Second International
  Conference on Knowledge Discovery \& Data Mining: Proceedings

\bibitem[\protect\citeauthoryear{Feldmann}{Feldmann}{2015}]{Feldmann2015}
Feldmann R.,  2015, \mn@doi [\mnras] {10.1093/mnras/stv552}, 449, 3274

\bibitem[\protect\citeauthoryear{{Ferrarotti} \& {Gail}}{{Ferrarotti} \&
  {Gail}}{2006}]{Ferrarotti2006}
{Ferrarotti} A.~S.,  {Gail} H.~P.,  2006, \mn@doi [\aap]
  {10.1051/0004-6361:20041198}, \href
  {https://ui.adsabs.harvard.edu/\#abs/2006A&A...447..553F} {447, 553}

\bibitem[\protect\citeauthoryear{Fiorentin, Bailer-Jones, Lee, Beers, Sivarani,
  Wilhelm, Prieto  \& Norris}{Fiorentin et~al.}{2007}]{Fiorentin2007}
Fiorentin P.~R.,  Bailer-Jones C. A.~L.,  Lee Y.~S.,  Beers T.~C.,  Sivarani
  T.,  Wilhelm R.,  Prieto C.~A.,   Norris J.~E.,  2007, \mn@doi [Astronomy
  {\&} Astrophysics] {10.1051/0004-6361:20077334}, 467, 1373

\bibitem[\protect\citeauthoryear{Fontanot \& Somerville}{Fontanot \&
  Somerville}{2011}]{Fontanot2011}
Fontanot F.,  Somerville R.~S.,  2011, \mn@doi [\mnras]
  {10.1111/j.1365-2966.2011.19245.x}, 416, 2962

\bibitem[\protect\citeauthoryear{{Galametz}, {Madden}, {Galliano}, {Hony},
  {Bendo}  \& {Sauvage}}{{Galametz} et~al.}{2011}]{Galametz2011}
{Galametz} M.,  {Madden} S.~C.,  {Galliano} F.,  {Hony} S.,  {Bendo} G.~J.,
  {Sauvage} M.,  2011, \mn@doi [\aap] {10.1051/0004-6361/201014904}, \href
  {http://adsabs.harvard.edu/abs/2011A%26A...532A..56G} {532, A56}

\bibitem[\protect\citeauthoryear{{Galliano}, {Madden}, {Jones}, {Wilson}  \&
  {Bernard}}{{Galliano} et~al.}{2005}]{Galliano2005}
{Galliano} F.,  {Madden} S.~C.,  {Jones} A.~P.,  {Wilson} C.~D.,   {Bernard}
  J.-P.,  2005, \mn@doi [\aap] {10.1051/0004-6361:20042369}, \href
  {http://adsabs.harvard.edu/abs/2005A%26A...434..867G} {434, 867}

\bibitem[\protect\citeauthoryear{Gehrz}{Gehrz}{1989}]{Gehrz1989}
Gehrz R.,  1989, in {Allamandola} L.~J.,  {Tielens} A.~G.~G.~M.,  eds,  IAU
  Symposium Vol. 135, Interstellar Dust. p.~445, \url
  {http://adsabs.harvard.edu/abs/1989IAUS..135..445G}

\bibitem[\protect\citeauthoryear{Gerdes, Sypniewski, McKay, Hao, Weis, Wechsler
   \& Busha}{Gerdes et~al.}{2010}]{Gerdes2010}
Gerdes D.~W.,  Sypniewski A.~J.,  McKay T.~A.,  Hao J.,  Weis M.~R.,  Wechsler
  R.~H.,   Busha M.~T.,  2010, \mn@doi [\apj] {10.1088/0004-637X/715/2/823},
  715, 823

\bibitem[\protect\citeauthoryear{Geurts, Ernst  \& Wehenkel}{Geurts
  et~al.}{2006}]{Geurts2006}
Geurts P.,  Ernst D.,   Wehenkel L.,  2006, \mn@doi [Machine Learning]
  {10.1007/s10994-006-6226-1}, 63, 3

\bibitem[\protect\citeauthoryear{Giannetti et~al.,}{Giannetti
  et~al.}{2017}]{Giannetti2017}
Giannetti A.,  et~al., 2017, \mn@doi [Astronomy {\&} Astrophysics]
  {10.1051/0004-6361/201731728}, 606, L12

\bibitem[\protect\citeauthoryear{{Gioannini}, {Matteucci}, {Vladilo}  \&
  {Calura}}{{Gioannini} et~al.}{2017}]{Gioannini2017}
{Gioannini} L.,  {Matteucci} F.,  {Vladilo} G.,   {Calura} F.,  2017, \mn@doi
  [\mnras] {10.1093/mnras/stw2343}, \href
  {https://ui.adsabs.harvard.edu/\#abs/2017MNRAS.464..985G} {464, 985}

\bibitem[\protect\citeauthoryear{{Goldsmith}}{{Goldsmith}}{2001}]{goldsmith01a}
{Goldsmith} P.~F.,  2001, \mn@doi [\apj] {10.1086/322255}, \href
  {http://adsabs.harvard.edu/abs/2001ApJ...557..736G} {557, 736}

\bibitem[\protect\citeauthoryear{Gong, Ostriker  \& Wolfire}{Gong
  et~al.}{2017}]{Gong2017}
Gong M.,  Ostriker E.~C.,   Wolfire M.~G.,  2017, \mn@doi [The Astrophysical
  Journal] {10.3847/1538-4357/aa7561}, 843, 38

\bibitem[\protect\citeauthoryear{Granato, Lacey, Silva, Bressan, Baugh, Cole
  \& Frenk}{Granato et~al.}{2000}]{Granato2000}
Granato G.~L.,  Lacey C.~G.,  Silva L.,  Bressan A.,  Baugh C.~M.,  Cole S.,
  Frenk C.~S.,  2000, \mn@doi [\apj] {10.1086/317032}, 542, 710

\bibitem[\protect\citeauthoryear{Hayward, Narayanan, Kere{\v s}, Jonsson,
  Hopkins, Cox  \& Hernquist}{Hayward et~al.}{2013}]{Hayward2013}
Hayward C.~C.,  Narayanan D.,  Kere{\v s} D.,  Jonsson P.,  Hopkins P.~F.,  Cox
  T.~J.,   Hernquist L.,  2013, \mn@doi [\mnras] {10.1093/mnras/sts222}, 428,
  2529

\bibitem[\protect\citeauthoryear{{Hirashita}}{{Hirashita}}{2000}]{Hirashita2000}
{Hirashita} H.,  2000, \mn@doi [Publications of the Astronomical Society of
  Japan] {10.1093/pasj/52.4.585}, \href
  {https://ui.adsabs.harvard.edu/\#abs/2000PASJ...52..585H} {52, 585}

\bibitem[\protect\citeauthoryear{Hirashita \& Kuo}{Hirashita \&
  Kuo}{2011}]{Hirashita2011}
Hirashita H.,  Kuo T.-M.,  2011, \mn@doi [\mnras]
  {10.1111/j.1365-2966.2011.19131.x}, 416, 1340

\bibitem[\protect\citeauthoryear{Hirashita \& Yan}{Hirashita \&
  Yan}{2009}]{Hirashita2009}
Hirashita H.,  Yan H.,  2009, \mn@doi [\mnras]
  {10.1111/j.1365-2966.2009.14405.x}, 394, 1061

\bibitem[\protect\citeauthoryear{Hirashita, Tajiri  \& Kamaya}{Hirashita
  et~al.}{2002}]{Hirashita2002}
Hirashita H.,  Tajiri Y.~Y.,   Kamaya H.,  2002, \mn@doi [Astronomy {\&}
  Astrophysics] {10.1051/0004-6361:20020605}, 388, 439

\bibitem[\protect\citeauthoryear{Hollenbach \& Salpeter}{Hollenbach \&
  Salpeter}{1971}]{Hollenbach1971}
Hollenbach D.,  Salpeter E.~E.,  1971, \mn@doi [\apj] {10.1086/150754}, 163,
  155

\bibitem[\protect\citeauthoryear{Hollenbach, Kaufman, Neufeld, Wolfire  \&
  Goicoechea}{Hollenbach et~al.}{2012}]{Hollenbach2012}
Hollenbach D.,  Kaufman M.~J.,  Neufeld D.,  Wolfire M.,   Goicoechea J.~R.,
  2012, \mn@doi [\apj] {10.1088/0004-637X/754/2/105}, 754, 105

\bibitem[\protect\citeauthoryear{Hopkins}{Hopkins}{2015}]{Hopkins2015}
Hopkins P.~F.,  2015, \mn@doi [\mnras] {10.1093/mnras/stv195}, 450, 53

\bibitem[\protect\citeauthoryear{{Hopkins} \& {Squire}}{{Hopkins} \&
  {Squire}}{2018a}]{Hopkins2018a}
{Hopkins} P.~F.,  {Squire} J.,  2018a, \mn@doi [\mnras]
  {10.1093/mnras/sty1604}, \href
  {https://ui.adsabs.harvard.edu/abs/2018MNRAS.479.4681H} {479, 4681}

\bibitem[\protect\citeauthoryear{{Hopkins} \& {Squire}}{{Hopkins} \&
  {Squire}}{2018b}]{Hopkins2018b}
{Hopkins} P.~F.,  {Squire} J.,  2018b, \mn@doi [\mnras]
  {10.1093/mnras/sty1982}, \href
  {https://ui.adsabs.harvard.edu/abs/2018MNRAS.480.2813H} {480, 2813}

\bibitem[\protect\citeauthoryear{Hopkins, Kere{\v s}, O{\~n}orbe,
  Faucher-Gigu{\`e}re, Quataert, Murray  \& Bullock}{Hopkins
  et~al.}{2014}]{Hopkins2014}
Hopkins P.~F.,  Kere{\v s} D.,  O{\~n}orbe J.,  Faucher-Gigu{\`e}re C.-A.,
  Quataert E.,  Murray N.,   Bullock J.~S.,  2014, \mn@doi [\mnras]
  {10.1093/mnras/stu1738}, 445, 581

\bibitem[\protect\citeauthoryear{Hopkins et~al.,}{Hopkins
  et~al.}{2018}]{Hopkins2018}
Hopkins P.~F.,  et~al., 2018, \mn@doi [\mnras] {10.1093/mnras/sty1690}, 480,
  800

\bibitem[\protect\citeauthoryear{Hou, Aoyama, Hirashita, Nagamine  \&
  Shimizu}{Hou et~al.}{2019}]{Hou2019}
Hou K.-C.,  Aoyama S.,  Hirashita H.,  Nagamine K.,   Shimizu I.,  2019,
  \mn@doi [\mnras] {10.1093/mnras/stz121}, 485, 1727

\bibitem[\protect\citeauthoryear{Inoue}{Inoue}{2003}]{Inoue2003}
Inoue A.~K.,  2003, \mn@doi [\pasj] {10.1093/pasj/55.5.901}, 55, 901

\bibitem[\protect\citeauthoryear{{Issa}, {MacLaren}  \& {Wolfendale}}{{Issa}
  et~al.}{1990}]{Issa1990}
{Issa} M.~R.,  {MacLaren} I.,   {Wolfendale} A.~W.,  1990, \aap, \href
  {https://ui.adsabs.harvard.edu/abs/1990A&A...236..237I} {236, 237}

\bibitem[\protect\citeauthoryear{Iwamoto, Brachwitz, Nomoto, Kishimoto, Umeda,
  Hix  \& Thielemann}{Iwamoto et~al.}{1999}]{Iwamoto1999}
Iwamoto K.,  Brachwitz F.,  Nomoto K.,  Kishimoto N.,  Umeda H.,  Hix W.~R.,
  Thielemann F.-K.,  1999, \mn@doi [\apjs] {10.1086/313278}, 125, 439

\bibitem[\protect\citeauthoryear{Jones, Tielens  \& Hollenbach}{Jones
  et~al.}{1996}]{Jones1996}
Jones A.~P.,  Tielens A. G. G.~M.,   Hollenbach D.~J.,  1996, \mn@doi [\apj]
  {10.1086/177823}, 469, 740

\bibitem[\protect\citeauthoryear{{Kahre} et~al.,}{{Kahre}
  et~al.}{2018}]{Kahre2018}
{Kahre} L.,  et~al., 2018, \mn@doi [\apj] {10.3847/1538-4357/aab101}, \href
  {https://ui.adsabs.harvard.edu/\#abs/2018ApJ...855..133K} {855, 133}

\bibitem[\protect\citeauthoryear{Kamdar, Turk  \& Brunner}{Kamdar
  et~al.}{2016}]{Kamdar2016}
Kamdar H.~M.,  Turk M.~J.,   Brunner R.~J.,  2016, \mn@doi [\mnras]
  {10.1093/mnras/stv2981}, 457, 1162

\bibitem[\protect\citeauthoryear{Katz, Laporte, Ellis, Devriendt  \& Slyz}{Katz
  et~al.}{2019}]{Katz2019}
Katz H.,  Laporte N.,  Ellis R.~S.,  Devriendt J.,   Slyz A.,  2019, \mn@doi
  [\mnras] {10.1093/mnras/stz281}, 484, 4054

\bibitem[\protect\citeauthoryear{Kennicutt}{Kennicutt}{1998}]{Kennicutt1998}
Kennicutt Jr. R.~C.,  1998, \mn@doi [\apj] {10.1086/305588}, 498, 541

\bibitem[\protect\citeauthoryear{Krumholz, McKee  \& Tumlinson}{Krumholz
  et~al.}{2009}]{Krumholz2009}
Krumholz M.~R.,  McKee C.~F.,   Tumlinson J.,  2009, \mn@doi [\apj]
  {10.1088/0004-637X/699/1/850}, 699, 850

\bibitem[\protect\citeauthoryear{{Krumholz}, {Leroy}  \& {McKee}}{{Krumholz}
  et~al.}{2011}]{krumholz11a}
{Krumholz} M.~R.,  {Leroy} A.~K.,   {McKee} C.~F.,  2011, \mn@doi [\apj]
  {10.1088/0004-637X/731/1/25}, \href
  {http://adsabs.harvard.edu/abs/2011ApJ...731...25K} {731, 25}

\bibitem[\protect\citeauthoryear{Lacey, Baugh, Frenk, Benson, Orsi, Silva,
  Granato  \& Bressan}{Lacey et~al.}{2010}]{Lacey2010}
Lacey C.~G.,  Baugh C.~M.,  Frenk C.~S.,  Benson A.~J.,  Orsi A.,  Silva L.,
  Granato G.~L.,   Bressan A.,  2010, \mn@doi [\mnras]
  {10.1111/j.1365-2966.2010.16463.x}, 405, 2

\bibitem[\protect\citeauthoryear{Lisenfeld \& Ferrara}{Lisenfeld \&
  Ferrara}{1998}]{Lisenfeld1998}
Lisenfeld U.,  Ferrara A.,  1998, \mn@doi [\apj] {10.1086/305354}, 496, 145

\bibitem[\protect\citeauthoryear{Ma et~al.,}{Ma et~al.}{2019}]{Ma2019}
Ma X.,  et~al., 2019, arXiv e-prints

\bibitem[\protect\citeauthoryear{{Magdis} et~al.,}{{Magdis}
  et~al.}{2012}]{Magdis2012}
{Magdis} G.~E.,  et~al., 2012, \mn@doi [\apj] {10.1088/0004-637X/760/1/6},
  \href {http://adsabs.harvard.edu/abs/2012ApJ...760....6M} {760, 6}

\bibitem[\protect\citeauthoryear{{Mathis}}{{Mathis}}{1990}]{Mathis1990}
{Mathis} J.~S.,  1990, \mn@doi [Annual Review of Astronomy and Astrophysics]
  {10.1146/annurev.aa.28.090190.000345}, 28, 37

\bibitem[\protect\citeauthoryear{{McKee}}{{McKee}}{1989}]{McKee1989}
{McKee} C.,  1989, in {Allamandola} L.~J.,  {Tielens} A.~G.~G.~M.,  eds,  IAU
  Symposium Vol. 135, Interstellar Dust. p.~431

\bibitem[\protect\citeauthoryear{{McKee}, {Hollenbach}, {Seab}  \&
  {Tielens}}{{McKee} et~al.}{1987}]{McKee1987}
{McKee} C.~F.,  {Hollenbach} D.~J.,  {Seab} G.~C.,   {Tielens} A.~G.~G.~M.,
  1987, \mn@doi [\apj] {10.1086/165403}, \href
  {https://ui.adsabs.harvard.edu/\#abs/1987ApJ...318..674M} {318, 674}

\bibitem[\protect\citeauthoryear{{McKinnon}, {Torrey}  \&
  {Vogelsberger}}{{McKinnon} et~al.}{2016}]{McKinnon2016}
{McKinnon} R.,  {Torrey} P.,   {Vogelsberger} M.,  2016, \mn@doi [\mnras]
  {10.1093/mnras/stw253}, \href
  {https://ui.adsabs.harvard.edu/\#abs/2016MNRAS.457.3775M} {457, 3775}

\bibitem[\protect\citeauthoryear{{McKinnon}, {Torrey}, {Vogelsberger},
  {Hayward}  \& {Marinacci}}{{McKinnon} et~al.}{2017}]{McKinnon2017}
{McKinnon} R.,  {Torrey} P.,  {Vogelsberger} M.,  {Hayward} C.~C.,
  {Marinacci} F.,  2017, \mn@doi [\mnras] {10.1093/mnras/stx467}, \href
  {http://adsabs.harvard.edu/abs/2017MNRAS.468.1505M} {468, 1505}

\bibitem[\protect\citeauthoryear{McKinnon, Vogelsberger, Torrey, Marinacci  \&
  Kannan}{McKinnon et~al.}{2018}]{McKinnon2018}
McKinnon R.,  Vogelsberger M.,  Torrey P.,  Marinacci F.,   Kannan R.,  2018,
  \mn@doi [\mnras] {10.1093/mnras/sty1248}, 478, 2851

\bibitem[\protect\citeauthoryear{M{\'e}nard, Kilbinger  \& Scranton}{M{\'e}nard
  et~al.}{2010}]{Menard2010}
M{\'e}nard B.,  Kilbinger M.,   Scranton R.,  2010, \mn@doi [\mnras]
  {10.1111/j.1365-2966.2010.16464.x}, 406, 1815

\bibitem[\protect\citeauthoryear{Morgan \& Edmunds}{Morgan \&
  Edmunds}{2003}]{Morgan2003}
Morgan H.~L.,  Edmunds M.~G.,  2003, \mn@doi [\mnras]
  {10.1046/j.1365-8711.2003.06681.x}, 343, 427

\bibitem[\protect\citeauthoryear{Muratov, Kere{\v s}, Faucher-Gigu{\`e}re,
  Hopkins, Quataert  \& Murray}{Muratov et~al.}{2015}]{Muratov2015}
Muratov A.~L.,  Kere{\v s} D.,  Faucher-Gigu{\`e}re C.-A.,  Hopkins P.~F.,
  Quataert E.,   Murray N.,  2015, \mn@doi [\mnras] {10.1093/mnras/stv2126},
  454, 2691

\bibitem[\protect\citeauthoryear{Nanni, Bressan, Marigo  \& Girardi}{Nanni
  et~al.}{2013}]{Nanni2013}
Nanni A.,  Bressan A.,  Marigo P.,   Girardi L.,  2013, \mn@doi [\mnras]
  {10.1093/mnras/stt1175}, 434, 2390

\bibitem[\protect\citeauthoryear{{Narayanan} \& {Dav{\'e}}}{{Narayanan} \&
  {Dav{\'e}}}{2012}]{narayanan12b}
{Narayanan} D.,  {Dav{\'e}} R.,  2012, \mn@doi [\mnras]
  {10.1111/j.1365-2966.2012.21159.x}, \href
  {http://adsabs.harvard.edu/abs/2012MNRAS.423.3601N} {423, 3601}

\bibitem[\protect\citeauthoryear{Narayanan et~al.,}{Narayanan
  et~al.}{2010}]{Narayanan2010}
Narayanan D.,  et~al., 2010, \mn@doi [\mnras]
  {10.1111/j.1365-2966.2010.16997.x}, 407, 1701

\bibitem[\protect\citeauthoryear{{Narayanan}, {Krumholz}, {Ostriker}  \&
  {Hernquist}}{{Narayanan} et~al.}{2011}]{narayanan11b}
{Narayanan} D.,  {Krumholz} M.,  {Ostriker} E.~C.,   {Hernquist} L.,  2011,
  \mn@doi [\mnras] {10.1111/j.1365-2966.2011.19516.x}, \href
  {http://adsabs.harvard.edu/abs/2011MNRAS.418..664N} {418, 664}

\bibitem[\protect\citeauthoryear{{Narayanan}, {Krumholz}, {Ostriker}  \&
  {Hernquist}}{{Narayanan} et~al.}{2012}]{narayanan12a}
{Narayanan} D.,  {Krumholz} M.~R.,  {Ostriker} E.~C.,   {Hernquist} L.,  2012,
  \mn@doi [\mnras] {10.1111/j.1365-2966.2012.20536.x}, \href
  {http://adsabs.harvard.edu/abs/2012MNRAS.421.3127N} {421, 3127}

\bibitem[\protect\citeauthoryear{{Narayanan} et~al.,}{{Narayanan}
  et~al.}{2015}]{Narayanan2015}
{Narayanan} D.,  et~al., 2015, \mn@doi [\nat] {10.1038/nature15383}, \href
  {http://adsabs.harvard.edu/abs/2015Natur.525..496N} {525, 496}

\bibitem[\protect\citeauthoryear{{Narayanan}, {Dav{\'e}}, {Johnson},
  {Thompson}, {Conroy}  \& {Geach}}{{Narayanan} et~al.}{2018a}]{narayanan18a}
{Narayanan} D.,  {Dav{\'e}} R.,  {Johnson} B.~D.,  {Thompson} R.,  {Conroy} C.,
    {Geach} J.,  2018a, \mn@doi [\mnras] {10.1093/mnras/stx2860}, \href
  {http://adsabs.harvard.edu/abs/2018MNRAS.474.1718N} {474, 1718}

\bibitem[\protect\citeauthoryear{{Narayanan}, {Conroy}, {Dav{\'e}}, {Johnson}
  \& {Popping}}{{Narayanan} et~al.}{2018b}]{narayanan18b}
{Narayanan} D.,  {Conroy} C.,  {Dav{\'e}} R.,  {Johnson} B.~D.,   {Popping} G.,
   2018b, \mn@doi [\apj] {10.3847/1538-4357/aaed25}, \href
  {http://adsabs.harvard.edu/abs/2018ApJ...869...70N} {869, 70}

\bibitem[\protect\citeauthoryear{{Ness}, {Hogg}, {Rix}, {Ho}  \&
  {Zasowski}}{{Ness} et~al.}{2015}]{Ness2015}
{Ness} M.,  {Hogg} D.~W.,  {Rix} H.~W.,  {Ho} A. Y.~Q.,   {Zasowski} G.,  2015,
  \mn@doi [\apj] {10.1088/0004-637X/808/1/16}, \href
  {https://ui.adsabs.harvard.edu/\#abs/2015ApJ...808...16N} {808, 16}

\bibitem[\protect\citeauthoryear{Niemi, Somerville, Ferguson, Huang, Lotz  \&
  Koekemoer}{Niemi et~al.}{2012}]{Niemi2012}
Niemi S.-M.,  Somerville R.~S.,  Ferguson H.~C.,  Huang K.-H.,  Lotz J.,
  Koekemoer A.~M.,  2012, \mn@doi [\mnras] {10.1111/j.1365-2966.2012.20425.x},
  421, 1539

\bibitem[\protect\citeauthoryear{Nomoto, Tominaga, Umeda, Kobayashi  \&
  Maeda}{Nomoto et~al.}{2006}]{Nomoto2006}
Nomoto K.,  Tominaga N.,  Umeda H.,  Kobayashi C.,   Maeda K.,  2006, \mn@doi
  [Nuclear Physics A] {10.1016/j.nuclphysa.2006.05.008}, 777, 424

\bibitem[\protect\citeauthoryear{Nozawa, Kozasa, Umeda, Maeda  \&
  Nomoto}{Nozawa et~al.}{2003}]{Nozawa2003}
Nozawa T.,  Kozasa T.,  Umeda H.,  Maeda K.,   Nomoto K.,  2003, \mn@doi [\apj]
  {10.1086/379011}, 598, 785

\bibitem[\protect\citeauthoryear{Nozawa, Kozasa, Habe, Dwek, Umeda, Tominaga,
  Maeda  \& Nomoto}{Nozawa et~al.}{2007}]{Nozawa2007}
Nozawa T.,  Kozasa T.,  Habe A.,  Dwek E.,  Umeda H.,  Tominaga N.,  Maeda K.,
   Nomoto K.,  2007, \mn@doi [\apj] {10.1086/520621}, 666, 955

\bibitem[\protect\citeauthoryear{{Nozawa}, {Maeda}, {Kozasa}, {Tanaka},
  {Nomoto}  \& {Umeda}}{{Nozawa} et~al.}{2011}]{Nozawa2011}
{Nozawa} T.,  {Maeda} K.,  {Kozasa} T.,  {Tanaka} M.,  {Nomoto} K.,   {Umeda}
  H.,  2011, \mn@doi [\apj] {10.1088/0004-637X/736/1/45}, \href
  {https://ui.adsabs.harvard.edu/\#abs/2011ApJ...736...45N} {736, 45}

\bibitem[\protect\citeauthoryear{Oppenheimer \& Dav{\'e}}{Oppenheimer \&
  Dav{\'e}}{2006}]{Oppenheimer2006}
Oppenheimer B.~D.,  Dav{\'e} R.,  2006, \mn@doi [\mnras]
  {10.1111/j.1365-2966.2006.10989.x}, 373, 1265

\bibitem[\protect\citeauthoryear{Ostriker \& Silk}{Ostriker \&
  Silk}{1973}]{Ostriker1973}
Ostriker J.,  Silk J.,  1973, \mn@doi [\apjl] {10.1086/181301}, 184, L113

\bibitem[\protect\citeauthoryear{Pedregosa et~al.,}{Pedregosa
  et~al.}{2012}]{Pedregosa2011}
Pedregosa F.,  et~al., 2012, Journal of Machine Learning Research, 12

\bibitem[\protect\citeauthoryear{Peek, M{\'e}nard  \& Corrales}{Peek
  et~al.}{2015}]{Peek2015}
Peek J. E.~G.,  M{\'e}nard B.,   Corrales L.,  2015, \mn@doi [\apj]
  {10.1088/0004-637X/813/1/7}, 813, 7

\bibitem[\protect\citeauthoryear{Peeples, Werk, Tumlinson, Oppenheimer,
  Prochaska, Katz  \& Weinberg}{Peeples et~al.}{2014}]{Peeples2014}
Peeples M.~S.,  Werk J.~K.,  Tumlinson J.,  Oppenheimer B.~D.,  Prochaska
  J.~X.,  Katz N.,   Weinberg D.~H.,  2014, \mn@doi [\apj]
  {10.1088/0004-637X/786/1/54}, 786, 54

\bibitem[\protect\citeauthoryear{{Pettini} \& {Pagel}}{{Pettini} \&
  {Pagel}}{2004}]{Pettini2004}
{Pettini} M.,  {Pagel} B. E.~J.,  2004, \mn@doi [\mnras]
  {10.1111/j.1365-2966.2004.07591.x}, \href
  {https://ui.adsabs.harvard.edu/abs/2004MNRAS.348L..59P} {348, L59}

\bibitem[\protect\citeauthoryear{{Pilyugin} \& {Grebel}}{{Pilyugin} \&
  {Grebel}}{2016}]{Pilyugin2016}
{Pilyugin} L.~S.,  {Grebel} E.~K.,  2016, \mn@doi [\mnras]
  {10.1093/mnras/stw238}, \href
  {https://ui.adsabs.harvard.edu/abs/2016MNRAS.457.3678P} {457, 3678}

\bibitem[\protect\citeauthoryear{{Planck Collaboration} et~al.,}{{Planck
  Collaboration} et~al.}{2016}]{Planck2016}
{Planck Collaboration} et~al., 2016, \mn@doi [Astronomy {\&} Astrophysics]
  {10.1051/0004-6361/201525830}, 594, A13

\bibitem[\protect\citeauthoryear{{Popping}, {Somerville}  \&
  {Galametz}}{{Popping} et~al.}{2017}]{Popping2017}
{Popping} G.,  {Somerville} R.~S.,   {Galametz} M.,  2017, \mn@doi [\mnras]
  {10.1093/mnras/stx1545}, \href
  {https://ui.adsabs.harvard.edu/\#abs/2017MNRAS.471.3152P} {471, 3152}

\bibitem[\protect\citeauthoryear{{Rafieferantsoa}, {Dav{\'e}}  \&
  {Naab}}{{Rafieferantsoa} et~al.}{2019}]{Rafieferantsoa2019}
{Rafieferantsoa} M.,  {Dav{\'e}} R.,   {Naab} T.,  2019, \mn@doi [\mnras]
  {10.1093/mnras/stz1199}, \href
  {https://ui.adsabs.harvard.edu/abs/2019MNRAS.486.5184R} {486, 5184}

\bibitem[\protect\citeauthoryear{{R{\'e}my-Ruyer} et~al.,}{{R{\'e}my-Ruyer}
  et~al.}{2014}]{Remy-Ruyer2014}
{R{\'e}my-Ruyer} A.,  et~al., 2014, \mn@doi [\aap]
  {10.1051/0004-6361/201322803}, \href
  {http://adsabs.harvard.edu/abs/2014A%26A...563A..31R} {563, A31}

\bibitem[\protect\citeauthoryear{{Rowlands}, {Gomez}, {Dunne},
  {Arag{\'o}n-Salamanca}, {Dye}, {Maddox}, {da Cunha}  \& {van der
  Werf}}{{Rowlands} et~al.}{2014}]{Rowlands2014}
{Rowlands} K.,  {Gomez} H.~L.,  {Dunne} L.,  {Arag{\'o}n-Salamanca} A.,  {Dye}
  S.,  {Maddox} S.,  {da Cunha} E.,   {van der Werf} P.,  2014, \mn@doi
  [\mnras] {10.1093/mnras/stu605}, \href
  {https://ui.adsabs.harvard.edu/abs/2014MNRAS.441.1040R} {441, 1040}

\bibitem[\protect\citeauthoryear{Schneider, Valiante, Ventura, dell'Agli,
  Di~Criscienzo, Hirashita  \& Kemper}{Schneider et~al.}{2014}]{Schneider2014}
Schneider R.,  Valiante R.,  Ventura P.,  dell'Agli F.,  Di~Criscienzo M.,
  Hirashita H.,   Kemper F.,  2014, \mn@doi [\mnras] {10.1093/mnras/stu861},
  442, 1440

\bibitem[\protect\citeauthoryear{{Seab} \& {Shull}}{{Seab} \&
  {Shull}}{1983}]{Seab1983}
{Seab} C.~G.,  {Shull} J.~M.,  1983, \mn@doi [\apj] {10.1086/161563}, \href
  {https://ui.adsabs.harvard.edu/\#abs/1983ApJ...275..652S} {275, 652}

\bibitem[\protect\citeauthoryear{Silva, Granato, Bressan  \& Danese}{Silva
  et~al.}{1998}]{Silva1998}
Silva L.,  Granato G.~L.,  Bressan A.,   Danese L.,  1998, \mn@doi [\apj]
  {10.1086/306476}, 509, 103

\bibitem[\protect\citeauthoryear{Smith et~al.,}{Smith et~al.}{2017}]{Smith2017}
Smith B.~D.,  et~al., 2017, \mn@doi [\mnras] {10.1093/mnras/stw3291}, 466, 2217

\bibitem[\protect\citeauthoryear{Somerville, Gilmore, Primack  \&
  Dom{\'{\i}}nguez}{Somerville et~al.}{2012}]{Somerville2012}
Somerville R.~S.,  Gilmore R.~C.,  Primack J.~R.,   Dom{\'{\i}}nguez A.,  2012,
  \mn@doi [\mnras] {10.1111/j.1365-2966.2012.20490.x}, 423, 1992

\bibitem[\protect\citeauthoryear{Sparre et~al.,}{Sparre
  et~al.}{2014}]{Sparre2014}
Sparre M.,  et~al., 2014, \mn@doi [\apj] {10.1088/0004-637X/785/2/150}, 785,
  150

\bibitem[\protect\citeauthoryear{Springel}{Springel}{2005}]{Springel2005}
Springel V.,  2005, \mn@doi [\mnras] {10.1111/j.1365-2966.2005.09655.x}, 364,
  1105

\bibitem[\protect\citeauthoryear{{Squire} \& {Hopkins}}{{Squire} \&
  {Hopkins}}{2018}]{Squire2018}
{Squire} J.,  {Hopkins} P.~F.,  2018, \mn@doi [\apj]
  {10.3847/2041-8213/aab54d}, \href
  {https://ui.adsabs.harvard.edu/abs/2018ApJ...856L..15S} {856, L15}

\bibitem[\protect\citeauthoryear{Temim, Dwek, Tchernyshyov, Boyer, Meixner,
  Gall  \& Roman-Duval}{Temim et~al.}{2015}]{Temim2015}
Temim T.,  Dwek E.,  Tchernyshyov K.,  Boyer M.~L.,  Meixner M.,  Gall C.,
  Roman-Duval J.,  2015, \mn@doi [\apj] {10.1088/0004-637X/799/2/158}, 799, 158

\bibitem[\protect\citeauthoryear{{Tielens}, {McKee}, {Seab}  \&
  {Hollenbach}}{{Tielens} et~al.}{1994}]{Tielens1994}
{Tielens} A.~G.~G.~M.,  {McKee} C.~F.,  {Seab} C.~G.,   {Hollenbach} D.~J.,
  1994, \mn@doi [\apj] {10.1086/174488}, \href
  {https://ui.adsabs.harvard.edu/\#abs/1994ApJ...431..321T} {431, 321}

\bibitem[\protect\citeauthoryear{Todini \& Ferrara}{Todini \&
  Ferrara}{2001}]{Todini2001}
Todini P.,  Ferrara A.,  2001, \mn@doi [\mnras]
  {10.1046/j.1365-8711.2001.04486.x}, 325, 726

\bibitem[\protect\citeauthoryear{Tremonti et~al.,}{Tremonti
  et~al.}{2004}]{Tremonti2004}
Tremonti C.~A.,  et~al., 2004, \mn@doi [\apj] {10.1086/423264}, 613, 898

\bibitem[\protect\citeauthoryear{{Tsai} \& {Mathews}}{{Tsai} \&
  {Mathews}}{1995}]{Tsai1995}
{Tsai} J.~C.,  {Mathews} W.~G.,  1995, \mn@doi [\apj] {10.1086/175943}, \href
  {https://ui.adsabs.harvard.edu/\#abs/1995ApJ...448...84T} {448, 84}

\bibitem[\protect\citeauthoryear{{Vijayan}, {Clay}, {Thomas}, {Yates},
  {Wilkins}  \& {Henriques}}{{Vijayan} et~al.}{2019}]{Vijayan2019}
{Vijayan} A.~P.,  {Clay} S.~J.,  {Thomas} P.~A.,  {Yates} R.~M.,  {Wilkins}
  S.~M.,   {Henriques} B.~M.,  2019, arXiv/1904.02196, \href
  {http://adsabs.harvard.edu/abs/2019arXiv190402196V} {}

\bibitem[\protect\citeauthoryear{Vlahakis, Dunne  \& Eales}{Vlahakis
  et~al.}{2005}]{Vlahakis2005}
Vlahakis C.,  Dunne L.,   Eales S.,  2005, \mn@doi [\mnras]
  {10.1111/j.1365-2966.2005.09666.x}, 364, 1253

\bibitem[\protect\citeauthoryear{Vogelsberger, McKinnon, O'Neil, Marinacci,
  Torrey  \& Kannan}{Vogelsberger et~al.}{2018}]{Vogelsberger2018}
Vogelsberger M.,  McKinnon R.,  O'Neil S.,  Marinacci F.,  Torrey P.,   Kannan
  R.,  2018, arXiv e-prints

\bibitem[\protect\citeauthoryear{Weingartner \& Draine}{Weingartner \&
  Draine}{2001}]{Weingartner2001}
Weingartner J.~C.,  Draine B.~T.,  2001, \mn@doi [\apj] {10.1086/324035}, 563,
  842

\bibitem[\protect\citeauthoryear{Wiseman, Schady, Bolmer, Krühler, Yates,
  Greiner  \& Fynbo}{Wiseman et~al.}{2017}]{Wiseman2017}
Wiseman P.,  Schady P.,  Bolmer J.,  Krühler T.,  Yates R.~M.,  Greiner J.,
  Fynbo J. P.~U.,  2017, \mn@doi [Astronomy {\&} Astrophysics]
  {10.1051/0004-6361/201629228}, 599, A24

\bibitem[\protect\citeauthoryear{Wolfire, Tielens, Hollenbach  \&
  Kaufman}{Wolfire et~al.}{2008}]{Wolfire2008}
Wolfire M.~G.,  Tielens A. G. G.~M.,  Hollenbach D.,   Kaufman M.~J.,  2008,
  \mn@doi [\apj] {10.1086/587688}, 680, 384

\bibitem[\protect\citeauthoryear{Yamasawa, Habe, Kozasa, Nozawa, Hirashita,
  Umeda  \& Nomoto}{Yamasawa et~al.}{2011}]{Yamasawa2011}
Yamasawa D.,  Habe A.,  Kozasa T.,  Nozawa T.,  Hirashita H.,  Umeda H.,
  Nomoto K.,  2011, \mn@doi [\apj] {10.1088/0004-637X/735/1/44}, 735, 44

\bibitem[\protect\citeauthoryear{Zafar \& Watson}{Zafar \&
  Watson}{2013}]{Zafar2013}
Zafar T.,  Watson D.,  2013, \mn@doi [Astronomy {\&} Astrophysics]
  {10.1051/0004-6361/201321413}, 560, A26

\bibitem[\protect\citeauthoryear{{Zhukovska}}{{Zhukovska}}{2014}]{Zhukovska2014}
{Zhukovska} S.,  2014, \mn@doi [\aap] {10.1051/0004-6361/201322989}, \href
  {http://adsabs.harvard.edu/abs/2014A%26A...562A..76Z} {562, A76}

\bibitem[\protect\citeauthoryear{Zhukovska, Gail  \& Trieloff}{Zhukovska
  et~al.}{2007}]{Zhukovska2007}
Zhukovska S.,  Gail H.-P.,   Trieloff M.,  2007, \mn@doi [Astronomy {\&}
  Astrophysics] {10.1051/0004-6361:20077789}, 479, 453

\makeatother
\end{thebibliography}

\bsp	
\label{lastpage}
\end{document}